\documentclass[10pt, letterpaper, twocolumn]{IEEEtran}

\usepackage{epsfig}
\usepackage{bm}
\usepackage[tbtags]{amsmath}
\usepackage{amsthm}
\usepackage{amssymb,amsfonts,latexsym,graphicx}
\usepackage{algorithmic}
\usepackage{subfigure}
\usepackage{graphics, psfrag}
\usepackage{enumerate}
\usepackage{epstopdf}        
\usepackage{cancel}
\usepackage{mathrsfs}
\usepackage{cite}
\usepackage{overpic}
\usepackage{color}

\graphicspath{{./figures/}}

\newcommand{\nnum}{\nonumber}

\newcommand{\ie}{i.e.}

\newtheorem{theorem}{Theorem}
\newtheorem{lemma}[theorem]{Lemma}

\newtheorem{proposition}[theorem]{Proposition}
\newtheorem{corollary}[theorem]{Corollary}
\newtheorem{definition}{Definition}

\newcommand{\ceil}[1]{\lceil{#1}\rceil}

\newcommand{\dotleq}{\stackrel{.}{\leq}}

\newcommand{\dotgeq}{\stackrel{.}{\geq}}

\newcommand{\eqa}{\stackrel{(a)}{=}}

\newcommand{\leb}{\stackrel{(b)}{\le}}

\DeclareMathOperator*{\argmax}{arg\,max}

\newcommand{\Hb}{H_{\mathrm{b}}} 

\newcommand{\calA}{\mathcal{A}}
\newcommand{\calB}{\mathcal{B}}
\newcommand{\calC}{\mathcal{C}}
\newcommand{\calD}{\mathcal{D}}
\newcommand{\calE}{\mathcal{E}}
\newcommand{\calF}{\mathcal{F}}

\newcommand{\calH}{\mathcal{H}}

\newcommand{\calK}{\mathcal{K}}

\newcommand{\calM}{\mathcal{M}}

\newcommand{\calR}{\mathcal{R}}
\newcommand{\calS}{\mathcal{S}}
\newcommand{\calT}{\mathcal{T}}
\newcommand{\calU}{\mathcal{U}}

\newcommand{\calW}{\mathcal{W}}
\newcommand{\calX}{\mathcal{X}}
\newcommand{\calY}{\mathcal{Y}}
\newcommand{\calZ}{\mathcal{Z}}

\newcommand{\bbN}{\mathbb{N}}

\newcommand{\bbR}{\mathbb{R}}

\newcommand{\scC}{\mathscr{C}}

\DeclareMathAlphabet{\mathbsf}{OT1}{cmss}{bx}{n}
\DeclareMathAlphabet{\mathssf}{OT1}{cmss}{m}{sl}

\newcommand{\rvA}{\mathsf{A}}

\newcommand{\rvB}{\mathsf{B}}

\newcommand{\rvE}{\mathsf{E}}

\newcommand{\rvP}{\mathsf{P}}

\DeclareSymbolFont{bsfletters}{OT1}{cmss}{bx}{n}
\DeclareSymbolFont{ssfletters}{OT1}{cmss}{m}{n}
\DeclareMathSymbol{\bsfGamma}{0}{bsfletters}{'000}
\DeclareMathSymbol{\ssfGamma}{0}{ssfletters}{'000}
\DeclareMathSymbol{\bsfDelta}{0}{bsfletters}{'001}
\DeclareMathSymbol{\ssfDelta}{0}{ssfletters}{'001}
\DeclareMathSymbol{\bsfTheta}{0}{bsfletters}{'002}
\DeclareMathSymbol{\ssfTheta}{0}{ssfletters}{'002}
\DeclareMathSymbol{\bsfLambda}{0}{bsfletters}{'003}
\DeclareMathSymbol{\ssfLambda}{0}{ssfletters}{'003}
\DeclareMathSymbol{\bsfXi}{0}{bsfletters}{'004}
\DeclareMathSymbol{\ssfXi}{0}{ssfletters}{'004}
\DeclareMathSymbol{\bsfPi}{0}{bsfletters}{'005}
\DeclareMathSymbol{\ssfPi}{0}{ssfletters}{'005}
\DeclareMathSymbol{\bsfSigma}{0}{bsfletters}{'006}
\DeclareMathSymbol{\ssfSigma}{0}{ssfletters}{'006}
\DeclareMathSymbol{\bsfUpsilon}{0}{bsfletters}{'007}
\DeclareMathSymbol{\ssfUpsilon}{0}{ssfletters}{'007}
\DeclareMathSymbol{\bsfPhi}{0}{bsfletters}{'010}
\DeclareMathSymbol{\ssfPhi}{0}{ssfletters}{'010}
\DeclareMathSymbol{\bsfPsi}{0}{bsfletters}{'011}
\DeclareMathSymbol{\ssfPsi}{0}{ssfletters}{'011}
\DeclareMathSymbol{\bsfOmega}{0}{bsfletters}{'012}
\DeclareMathSymbol{\ssfOmega}{0}{ssfletters}{'012}

\newcommand{\tilE}{\tilde{E}}

\newcommand{\tilF}{\tilde{F}}

\newcommand{\tilH}{\tilde{H}}

\newcommand{\hatm}{\hat{m}}
\newcommand{\hatM}{\hat{M}}
\newcommand{\tilm}{\tilde{m}}
\newcommand{\tilM}{\tilde{M}}

\newcommand{\tilq}{\tilde{q}}

\newcommand{\hatx}{\hat{x}}
\newcommand{\hatX}{\hat{X}}
\newcommand{\tilx}{\tilde{x}}
\newcommand{\tilX}{\tilde{X}}

\def\fndot{\, \cdot \,}

\newcommand{\fcost}{\Lambda}    

\newcommand{\KA}{K_{\mathrm{A}}}  
\newcommand{\KB}{K_{\mathrm{B}}}  
\newcommand{\kA}{k_{\mathrm{A}}}  
\newcommand{\kB}{k_{\mathrm{B}}}  

\newcommand{\xa}{X}		
\newcommand{\xb}{Y}
\newcommand{\xe}{Z}
\newcommand{\xin}{S}

\newcommand{\alpb}{\calY}
\newcommand{\alpe}{\calZ}

\newcommand{\ka}{K_{\mathrm{A}}}
\newcommand{\kb}{K_{\mathrm{B}}}
\newcommand{\pmsg}{\Phi}  
\newcommand{\alppmsg}{\varPhi}

\newcommand{\Rsk}{R_{\mathrm{SK}}}	
\newcommand{\Rpub}{R_{\Phi}}	
\newcommand{\Csk}{C_{\mathrm{SK}}}	

\newcommand{\randBern}[1]{\mathrm{Bern}\left(#1\right)}  

\newcommand{\Er}{E_{\mathrm{r}}}
\newcommand{\Fr}{F_{\mathrm{r}}}
\newcommand{\Eo}{E_{\mathrm{o}}}
\newcommand{\Fo}{F_{\mathrm{o}}}
\newcommand{\tilEo}{\tilE_{\mathrm{o}}}
\newcommand{\tilFo}{\tilF_{\mathrm{o}}}

\newcommand{\ind}{{\bf 1}}
\newcommand{\indVar}[4]{{\bf 1}[#3, #4|#1,#2, \scC]}
\newcommand{\indA}{\ind[k_{\mathrm{A}}=k'_{\mathrm{A}}, \phi=\phi']}
\newcommand{\indB}{\indVar{m'}{x'^n}{k'_{\mathrm{A}}}{\phi'}}
\newcommand{\indD}{\indVar{m}{x^n}{k_{\mathrm{A}}}{\phi}}
\newcommand{\indDD}{{\bf 1}[k_{\mathrm{A}}, \phi|m,x^n,\calC]}
\newcommand{\indE}{\indVar{m'}{x'^n}{k_{\mathrm{A}}}{\phi}}

\title{The Sender-Excited Secret Key Agreement Model: Capacity, Reliability and Secrecy Exponents}
\author{Tzu-Han Chou, \hspace{.025in} Vincent
  Y.~F.\ Tan, \hspace{.025in} Stark C. Draper \thanks{This work was
    supported in part by the Air Force Office of Scientific Research
    under grant FA9550-09-1-0140, by a grant from the Wisconsin Alumni
    Research Foundation, and by the National Science Foundation under
    CAREER grant CCF 0844539.  The work of V.~Y.~F.\ Tan was also
    supported by A*STAR, Singapore.  This paper was presented in part
    at Allerton Conference on Communication, Control and Computing in
    Monticello, IL (September 2011)~\cite{chou_11}.  }
  \thanks{T.-H.~Chou is with Qualcomm Inc, San Diego, CA. V.~Y.~F.~Tan
    is with the Institute of Infocomm Research, Singapore and the
    Department of Electrical and Computer Engineering, National
    University of Singapore. S.~C.~Draper is with the Department of
    Electrical and Computer Engineering, University of Wisconsin,
    Madison, WI, 53706, USA (emails: {tzuhanc@qti.qualcomm.com};
    {vtan@nus.edu.sg}; {sdraper@ece.wisc.edu}).  }}

\begin{document}
\maketitle

\begin{abstract}
We consider the secret key generation problem when sources are
randomly excited by the sender and there is a noiseless public
discussion channel.  Our setting is thus similar to recent works on
channels with action-dependent states where the channel state may be
influenced by some of the parties involved. We derive single-letter
expressions for the secret key capacity through a type of source
emulation analysis.  We also derive lower bounds on the achievable
reliability and secrecy exponents, i.e., the exponential rates of
decay of the probability of decoding error and of the information
leakage. These exponents allow us to determine a set of
strongly-achievable secret key rates.  For degraded eavesdroppers the
maximum strongly-achievable rate equals the secret key capacity; our
exponents can also be specialized to previously known results.

In deriving our strong achievability results we introduce a coding
scheme that combines wiretap coding (to excite the channel) and key
extraction (to distill keys from residual randomness).  The secret key
capacity is naturally seen to be a combination of both source- and
channel-type randomness.  Through examples we illustrate a fundamental
interplay between the portion of the secret key rate due to each type
of randomness.  We also illustrate inherent tradeoffs between the
achievable reliability and secrecy exponents.  Our new scheme also
naturally accommodates rate limits on the public discussion.  We show
that under rate constraints we are able to achieve larger rates than
those that can be attained through a pure source emulation strategy.

\end{abstract}


\section{Introduction}
\label{sec:introduction}

Within the realm of information-theoretic secrecy \cite{Liang}, the
foundations of sharing a secret key between two parties in the
presence of an eavesdropper were initiated
in~\cite{Ahlswede_Csiszar93, maurer93}.  Ahlswede and
Csisz\'{a}r~\cite{Ahlswede_Csiszar93} studied two models: the
\emph{source-type model with wiretapper} (Model SW) and the
\emph{channel-type model with wiretapper} (Model CW). In Model SW,
users obtain their observations from a discrete memoryless multiple
source (DMMS), and communicate to each other via a noiseless
authenticated public channel, with the objective of generating jointly
held secret keys.  In Model CW, one legitimate user (the sender)
controls the input of a discrete memoryless broadcast channel (DMBC),
sending information based upon which the legitimate receivers generate
secret keys.  


However, many applications cannot be exactly modeled as either a
source- or a channel-type scenario. This work explores such a setting
in which the sender has the ability to use a private source of
randomness to excite (or influence) the ``state'' of the DMMS. This is
similar in spirit to recent works on probing capacity and channels
with action-dependent
states~\cite{Weissman10,Asnani_ProbingCapacity,Kit10,Per11}.  We
derive capacity, reliability exponent, and secrecy exponent results
for this setting.  At one extreme, when the sender has an unlimited
ability to excite the channel, and the rate of public discussion is
similarly unbounded, a particular type of source emulation strategy is
capacity achieving.  However, when constraints are placed on the rate
of public discussion we demonstrate that source emulation becomes
sub-optimal.  We show this through the development of a more nuanced
rate-limited excitation strategy that exceeds the capacity of the
emulation-based approach when subject to rate
constraints~\cite{csiszar2000}.  Our new strategy combines a
wiretap-type probing mechanism (Model CW) with a key-distillation step
(Model SW) that is applied to the residual randomness.  In general, we
find an interplay to exist between the secrecy rate derived from the
wiretapping step and the secrecy rate derived via the key-distillation
step.  We illustrate the tradeoff via examples.  In terms of our large
deviation results we show that there is a natural tradeoff between the
reliability and secrecy exponents.  The former generalize Gallager's
classic results in in~\cite[Sec.\ 5.6]{gallagerIT}
and~\cite{gallager76}; the latter may be specialized to Hayashi's
recent work that characterizes the rate of decay of information
leakage~\cite{Hayashi} of the wiretap channels.

\subsection{Related Work}

There are other investigations that consider non-source, non-channel
models.  For example, in~\cite{Khisti_isit08, Prabhakaran08} users
observe a DMMS and can also transmit information via a wiretap
channel. However, no public discussion is allowed. The key generation
scheme used is based on the observation that a public message can be
transmitted via the DMBC confidentially, resulting in a higher secret
key rate. In~\cite{csiszar2000, CsiszarN04, CsiszarN08}, public
discussion is allowed and there may also be a helper.  However, unlike
our work, the sender does not also receive a sequence as part of the
channel output. The sender's ability to use both her channel output
and her source of private randomness to generate the secret key is a
crucial aspects of our model.

The authors in~\cite{ChenVinck06, Liu07, Chia10, Khisti_isit09,
  Khisti11} considered the setting where a wiretap channel is
influenced by a random state that is known by the sender (and possibly
by the receiver) and thus can be treated as a correlated source. In
\cite{ChenVinck06, Liu07}, the sender transmits a confidential message
and the random, noncausally known, state is exploited to confuse the
eavesdropper. The lower bound is proved using a combination of
Gel'fand-Pinsker coding and wiretap channel coding. A similar problem
but with causal state information is studied in \cite{Chia10} and the
coding scheme involves block Markov coding, Shannon strategies, and
wiretap coding. In~\cite{Khisti_isit09, Khisti11}, the goal is to
generate a secret key when the encoder (and/or decoders) have
noncausal state information. The authors present a single-letter
expression for the secret key capacity. The key rate consists of two
parts.  The first can be attributed to the rate of the confidential
message sent using wiretap channel coding where the state sequence is
treated as a time-sharing sequence, while a second key, independent of
the first, is produced by exploiting the common knowledge of the state
at the sender and the legitimate receiver.



%

The model considered in this paper is a generalization of the ``source
excitation'' model of~\cite{chou_it10}.  That model is motivated by
the large body of work on physical-layer security (see, e.g.,
\cite{WTS07, ARKA11}) where the unpredictable variation in the
wireless channel medium serves as the source of common randomness.
One approach is to sound the wireless channel using a random signal
and measure the observations generated (marginalizing over the
sounding signal).  This ``source emulation'' strategy is considered in
\cite{ARKA11}.  Another approach studied in \cite{chou_it10,WTS07}
uses deterministic sounding (no marginalization is involved).  Key
extraction follows by denoising the observations using a public
message.  Deterministic sounding requires no source of private
randomness (as does source emulation), all randomness is due to the
channel.  The current generalization is that we now explore the source
excitation model when the exciter has a source of private randomness.
This allows us to exploit both random sounding (using a wiretap code)
and key generation (using conditional randomness).  We regard the
current model as stepping stone to understanding the fundamental
limits of two-way randomized channel sounding in which secrecy rate is
derived from the use of two wiretap codes and from the conditional
randomness produced.

\subsection{Main Contributions: Capacity and Error Exponents}
\begin{figure}
\centering
\begin{picture}(200, 133)
\linethickness{.25mm}
\put(0,0){\line(1,0){200}}
\put(0,10){\line(1,0){200}}
\put(0,0){\line(0,1){10}}
\put(200,0){\line(0,1){10}}
\put(70,2){\mbox{Public Channel}}

\put(0,40){\line(1,0){50}}
\put(0,60){\line(1,0){50}}
\put(0,40){\line(0,1){20}}
\put(50,40){\line(0,1){20}}
\put(16,47){\mbox{Alice}}

\put(75,40){\line(1,0){50}}
\put(75,60){\line(1,0){50}}
\put(75,40){\line(0,1){20}}
\put(125,40){\line(0,1){20}}
\put(91,47){\mbox{Bob}}

\put(150,40){\line(1,0){50}}
\put(150,60){\line(1,0){50}}
\put(150,40){\line(0,1){20}}
\put(200,40){\line(0,1){20}}
\put(166,47){\mbox{Eve}}

\put(65,80){\line(1,0){70}}
\put(65,100){\line(1,0){70}}
\put(65,80){\line(0,1){20}}
\put(135,80){\line(0,1){20}}
\put(78,87){\mbox{$p(x,y,z|s)$}}

\put(25,60){\vector(0,1){50}}
\put(50,120){\line(1,0){50}}
\put(100,120){\vector(0,-1){20}}

\put(9,117){\mbox{Encoder}}

\put(0,110){\line(1,0){50}}
\put(0,130){\line(1,0){50}}
\put(0,110){\line(0,1){20}}
\put(50,110){\line(0,1){20}}

\put(85,110){\mbox{$S^n$}}

\put(13,85){\mbox{$M$}}

\put(100,80){\vector(0,-1){20}}
\put(105,67){\mbox{$Y^n$}}

\put(65,90){\line(-1,0){25}}
\put(40,90){\vector(0,-1){30}}
\put(43,67){\mbox{$X^n$}}

\put(135,90){\line(1,0){40}}
\put(175,90){\vector(0,-1){30}}
\put(177,67){\mbox{$Z^n$}}

\put(10,40){\vector(0,-1){30}}
\put(0,28){\mbox{$\Phi$}}
 
\put(40,40){\vector(0,-1){14}} 
\put(32,16){\mbox{$K_{\mathrm{A}}$}}

\put(115,40){\vector(0,-1){14}}
\put(110,16){\mbox{$K_{\mathrm{B}}$}}

\put(75,28){\mbox{$\Phi$}}
\put(85,10){\vector(0,1){30}}

\put(165,28){\mbox{$\Phi$}}
\put(175,10){\vector(0,1){30}}

\end{picture}
\caption{Our problem setup: Based on her private source  of randomness $M$, Alice excites the channel via the sounding signal $S^n(M)$. She generates a public message $\Phi(M, X^n)$, which is transmitted through the noiseless public channel and hence  known to all parties.   Alice and Bob generate keys $K_{\mathrm{A}}(M, X^n)$ and $K_{\mathrm{B}}(\Phi,Y^n)$ respectively. The keys should agree, while at the same time, they should be kept secret from Eve.}
\label{fig:sk}
\end{figure}
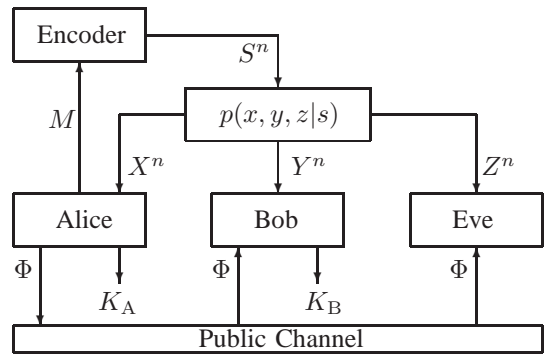

Figure~\ref{fig:sk} shows the system considered in this paper.  We can
think of the terminal labeled Alice as a base station on earth
equipped with a sensor. This base station transmits a random message
$M$ (the selection of which is based on a private source of
randomness) securely to a satellite encoder.  The satellite produces
sequence $S^n$ according to some conditional probability law.  This
sequence is the input to a broadcast channel $p(x,y,z|s)$ (the
wireless medium). The channel produces observations $X^n$, $Y^n$ and
$Z^n$, respectively received by Alice, the legitimate user Bob, and
the malicious user Eve.  The goal of the two legitimate users is to
generate a shared secret key -- Alice based on $(M,X^n)$ and Bob based
on $(\Phi,Y^n)$, where $\Phi$ is a public message known to all
parties.

We first consider the situation in which there are no rate limits on
either the public discussion ($\Phi$) or the excitation signal ($M$).
We derive a single-letter expression for the secret key capacity of
this system.  The result follows through a particular kind of source
emulation where (i) Alice chooses the optimum source distribution to
induce (potentially subject to cost constraints on $S^n$), and (ii)
Alice has the vector observation $(S^n, X^n)$.


We then turn to the rate-limited situation and study the effect of
rate limits on (i) the achievable secrecy rate, (ii) the probability
of erroneous decoding at the legitimate receiver, Bob, and (iii) the
key leakage rate by the eavesdropper, Eve. We focus on degraded
channels and characterize the error probability in terms of a
\emph{reliability exponent} and the key leakage rate in terms of a
\emph{secrecy exponent}.  In contrast to~\cite{csiszar2000} where the
secret key capacity of one-way key generation subject to a rate
constraint is characterized, we show that the flexibility Alice has in
choosing the amount of private randomness she uses in the selection of
$M$ can allow a strictly higher achievable secret key rate than can be
attained via pure source emulation.

We introduce a new type of decoder for the legitimate receiver, Bob,
to use.  This decoder is a combination of a maximum likelihood and a
maximum {\em a-posteriori} (ML-MAP) decoder.  Bob decodes jointly the
sender's source $X^n$ and the sender's private source of randomness
(or message) $M$. The resulting reliability exponent expression can be
specialized to Gallager's channel coding error
exponent~\cite[Sec.\ 5.6]{gallagerIT} and Gallager's source coding
error exponent~\cite{gallager76}. On the other hand, in the key
leakage analysis, the secrecy exponent we derive captures the leakage
due to Eve's channel $p(z|s)$ and the leakage due to the correlation
between Alice's variable $X$ and Eve's variable $Z$ in a transparent
manner. Our analysis builds on the work by Hayashi in~\cite{Hayashi,
  Hayashi06}, where he links the leakage rate of a wiretap channel to
channel resolvability and identification coding~\cite{Han}. This
connection is also examined Bloch and Laneman~\cite{Bloch} where they
derive the capacity of general wiretap channels from an information
spectrum perspective~\cite{Han}. Our secrecy exponent results, which
are developed in Section \ref{sec:exponent_result}, can be specialized
to the wiretap channel~\cite{Hayashi, Hayashi06} and to the secret key
generation from correlated source setting~\cite{BBCM95,Wat10, Hayashi,
  chou_it10}. The difference vis-\`{a}-vis the motivating
work~\cite{chou_it10} is that the methods used to bound the exponents
for both reliability and secrecy involve both wiretap channel coding
and source coding. This will become clear in
Section~\ref{sec:exponent_result} where we specialize our results to
various known problems. Note that the criterion for exponential decay
of the key leakage rate is much stronger than the usual strong secrecy
\cite{maurer93}. We focus on this exponential notion because it
quantifies how fast the error probability and information rate decays
to zero and because it reveals a natural tradeoff between the
attainable reliability and secrecy exponents.

\subsection{Paper Organization}
This paper is organized as follows: In
Section~\ref{sec:problem_setting}, we describe the system model. We
also define the secret key capacity, the capacity-reliability-secrecy
region and the notion of channel degradedness. Our main results
pertaining to the secret key capacity are provided in
Section~\ref{sec:capacity_result}. We also prove a (sometimes loose)
upper bound on the secret key capacity that does not contain any
auxiliary random variables, and hence is amenable to evaluation. We
show that this upper bound is tight for degraded channels. We present
the reliability and secrecy exponents in
Section~\ref{sec:exponent_result} and connect to previous work. In
Section~\ref{sec:example}, we present several examples to demonstrate
how the main results can be applied to channels of interest. We show
the inherent tradeoff between the portions of the secret key rate due
to source- and to channel-type randomness. We also show the inherent
tradeoff between the reliability exponent and the secrecy exponent.  The proofs of the capacity theorems and
the error exponent theorems are provided in
Section~\ref{sec:proof_capacity} and Section~\ref{sec:proof_exponent}
respectively.

\subsection{Notation}
We generally adopt the notational conventions in the book by El Gamal
and Kim~\cite{ElGamal_Kim_LNIT}, some of which we recap here. All
logarithms are to base-$2$. Random variables are in upper case (e.g.,
$X$) and their realizations in lower case (e.g., $x$). The
corresponding alphabets of random variables are in calligraphic font
(e.g., $\calX$) and so are all sets and events (e.g., $\scC$). For
vectors, $X^i_j \triangleq (X_j,\ldots, X_i)$ and if $j=1$, the
abbreviation $X^i \triangleq X^i_1$ is used. In addition,
$X^{n\setminus i } \triangleq (X^{i-1}, X_{i+1}^n)$.  The probability
mass function (pmf) of a discrete random variable $X$ is denoted as
$p_X(x)$ or more simply as $p(x)$.  Random codebooks are denoted by a
special script font $\scC$ while a codebook realization is denoted as
$\calC$.  For an $a\ge 0$, we also commonly use the notation
$[1:2^{a}]\triangleq\{1,\ldots, 2^{\ceil{a}}\}$.


\section{Problem Setup}
\label{sec:problem_setting}

\subsection{The Secret Key Generation Protocol} \label{sec:protocol}
The setting is shown in Fig.~\ref{fig:sk}.   Consider a 3-receiver DMBC $(\calS, p(x,y,z|s), \calX\times\calY\times \calZ)$ consisting of four finite sets $\calS,\calX,\calY,\calZ$ and a collection of conditional pmfs $p(x,y,z|s)$ on $\calX\times\calY\times \calZ$.   Alice, at terminal $\calX$,  controls the channel input  {\em sounding signal}  $s^n$ through the encoder via $n$ uses of the channel.   Alice has a private source of randomness used to select an index $m$, which influences $s^n$.   The legitimate receiver at terminal $\alpb$ is known as Bob  and  the eavesdropper at terminal $\alpe$  is known as Eve. There is also a noiseless public discussion channel which allows Alice to transmit  a message $\Phi$ to Bob and Eve.  Let $\Lambda:\calS\to [0,\Lambda_{\max}]$ be a per-letter, bounded cost function and let $\Gamma>0$ be an admissible cost.   A $(2^{nR_M}, 2^{nR_{\Phi}}, n,\Gamma)$ {\em code}  for the  secret key generation protocol  consists of a tuple of functions $(f,\phi,k_{\mathrm{A}})$. In particular, 
\begin{enumerate}
  \item \emph{Channel Excitation}: Alice        selects a  message $M \in [1:2^{nR_M}]$ uniformly at random.   The (satellite) encoder   sends a message-dependent input sequence  $S^n = f(M) \in\calS^n$ ($f$ possibly being random) satisfying
      \begin{equation}\label{equ:cost_constraint}
        \rvP\left[ \frac{1}{n} \sum_{i=1}^n \fcost(S_i) \le\Gamma  \right] = 1 \ .
      \end{equation}
     The  input sequence  $S^n$ is  transmitted over $n$ uses of $p(x,y,z|s)$. The output sequences $x^n$, $y^n$ and $z^n$ are observed by Alice, Bob (legitimate receiver) and Eve (eavesdropper)  respectively.

  \item \emph{One-Way (Forward) Public Discussion}: After observing $x^n$, Alice generates a one-way public  message\footnote{As in \cite{ElGamal_Kim_LNIT}, we use a common notation $\phi$ to denote both the function $\phi: [ 1:2^{nR_M}]\times\calX^n\to [1:2^{nR_{\pmsg}} ]$ as well as the output of the function $\phi \in [1:2^{nR_{\pmsg}} ]$. This applies in the rest of the paper.}  $\phi = \phi(m, x^n) \in [1:2^{nR_{\pmsg}} ]$, and transmits it over a noiseless public channel.
  \item \emph{Key Generation}: Alice generates a key $k_{\mathrm{A}} = k_{\mathrm{A}}(m, x^n) \in\bbN$. After receiving  his channel output $y^n$  and the public message $\phi$, Bob generates another key $k_{\mathrm{B}} = k_{\mathrm{B}}(y^n, \phi) \in\bbN$.
\end{enumerate}



Note the conditional distribution of $( \xa, \xb, \xe)$ given $\xin$
can be factorized as $p(x|s)p(y,z|x,s)$. The first conditional
distribution $p(x|s)$ can be roughly thought of as Alice's influence
on the channel state via the sounding signal $s^n$, while the second
$p(y,z|x,s)$ can be thought of as a state-dependent channel.



\subsection{Definitions}
We now provide the  definitions of achievable secret key rates, secret key capacity  and error exponents. As a reminder, the random variables $\KA$ and $\KB$ respectively denote Alice's and Bob's key. The public message is denoted as $\Phi$.

\begin{definition}[Weak  Achievability]  \label{def:wach} The  secret key  rate   $\Rsk  \in\bbR_+$   is   {\em $\Gamma$-weakly-achievable} (or simply $\Gamma$-achievable)  if there exists a sequence of  $(2^{nR_M}, 2^{nR_{\Phi}},  n,\Gamma)$ codes  (for any $(R_M,R_\Phi)$ pair)   for the  secret key generation protocol such that   the following  three  conditions   are satisfied: 
\begin{align}
      \lim_{n\to\infty}  \,\,\quad \; \rvP(\ka \neq \kb)   &= 0  \ , \label{eqn:agreement} \\
\lim_{n\to\infty}    \,\,\, \; \frac{1}{n} I(\ka ; \xe^n, \pmsg) &  =  0 \ ,  \label{equ:secrecy_condition}\\
\liminf_{n\to\infty}   \,\,\qquad \;   \frac{1}{n} H(\ka)  &  \ge  \Rsk  \ , \label{eqn:sk_rate}
\end{align}
\end{definition}

\begin{definition}[(Forward) Secret Key Capacity] \label{def:cap}
 The {\em secret key capacity-cost function} $\Csk(\Gamma)$ is defined as follows:
 \begin{equation}
\Csk(\Gamma) :=\sup\{\Rsk: \Rsk \mbox{ is } \Gamma\mbox{-weakly-achievable}\} \ .
\end{equation}
\end{definition}

We will henceforth say that $\Csk(\Gamma)$  is the {\em (forward) secret key capacity} (without reference to the cost $\Gamma$).   The {\em reliability condition} in~\eqref{eqn:agreement} implies that we would like Alice's and Bob's keys to agree with high probability. The {\em secrecy condition}  in~\eqref{equ:secrecy_condition} requires  that the eavesdropper cannot estimate the key $K_{\mathrm{A}} \in [1:2^{n R_{\mathrm{SK}}}]$ given   her  observation $Z^n$ and the  public message $\Phi$. This is manifested in  that the   {\em key leakage rate} $\frac{1}{n} I(\ka ; \xe^n, \pmsg)$ is arbitrarily small for sufficiently large blocklength $n$.   The rate   condition in~\eqref{eqn:sk_rate} implies that the entropy of $K_{\mathrm{A}}$ should be close to $\Rsk$. In other words the pmf of $K_{\mathrm{A}}$ should be close to that of a uniform pmf   on  $[1:2^{n\Rsk }]$, so the eavesdropper can only glean a negligible amount of information.

In many practical settings, the fact that the error probability in~\eqref{eqn:agreement}  and the key leakage rate in \eqref{equ:secrecy_condition} can be made arbitrarily small with increasing block length is insufficient. See Maurer's work in~\cite{Mau00} and a more recent exposition in~\cite{Bloch}. It would, in fact, be desirable to    quantify  their  rates of decay and to devise coding schemes to ensure that these decay rates are as large as possible. We  formalize this by defining the notion of an achievable secret key rate-exponent triple. To simplify the exposition, in our definitions (and corresponding results) of rates with exponents, we will  assume that $\Gamma=\infty$. In other words, we do not impose a  cost constraint on $S^n$ as in~\eqref{equ:cost_constraint}.  


\begin{definition}[Achievable Secret Key Rate-Exponent  Triple]\label{def:rate-expo-triple}
    The secret key rate-exponent triple $(\Rsk, E,F) \in \bbR_+^3$ is  {\em achievable} if there exists a sequence of $(2^{nR_M}, 2^{n\Rpub},   n)$  codes  for the  secret key generation protocol such that in addition to~\eqref{eqn:sk_rate}, the following hold:
    \begin{align}
        &\liminf_{n\to\infty}\,  \, -\frac{1}{n} \log \rvP(\KA\ne\KB) \geq E \ ,  \label{eqn:Eexponent}\\
        &\liminf_{n\to\infty}\,  \, -\frac{1}{n} \log I( \KA; \xe^n,\Phi)   \geq  F \ . \label{eqn:Fexponent}
    \end{align}
\end{definition}

In~\eqref{eqn:Eexponent}, $E$ is known as  the \emph{reliability exponent} and  in~\eqref{eqn:Fexponent}, $F$ is known as  the \emph{secrecy exponent}.  Collectively, $E$ and $F$ are known as {\em error exponents} (though    $ I( \KA; \xe^n,\Phi)$ is not, strictly speaking, an error probability but we abuse terminology to say that both are ``errors''). Definition \ref{def:rate-expo-triple} can also be interpreted as follows: If a triple $(\Rsk , E,F)$  is achievable, then the error probability in~\eqref{eqn:agreement}  decays\footnote{Here and in the following, for a pair of positive sequences $\{(a_n,b_n)\}_{n\in\bbN}$, we say that $a_n\dotleq b_n$ if $\limsup_{n\to\infty} n^{-1}\log (a_n/b_n)\le 0$. The notation $\dotgeq$ is defined analogously. We say that $a_n\doteq b_n$ if $a_n\dotleq b_n$ and $a_n\dotgeq b_n$.} as  $\rvP(\ka \neq \kb) \dotleq 2^{-nE}$ and the key leakage   decays as $I( \KA; \xe^n,\Phi)\dotleq 2^{-nF}$. Naturally, the constraint on the entropy of the secret key in~\eqref{eqn:sk_rate} is retained in the above definition.

\begin{definition}[Capacity-Reliability-Secrecy Region] \label{def:re_region}
The {\em (secret key)  capacity-reliability-secrecy region} $\calR \subset\bbR^3_+$ is the closure of the set of  achievable secret key rate-exponent triples.
\end{definition}

In analogy to the notion of weak achievability, we can also define a more stringent notion known as strong achievability, also studied in~\cite{Maurer94,Mau00}.
\begin{definition}[Strong Achievability]\label{def:strong_achievable}
    The secret key rate $\Rsk$ is \emph{strongly-achievable} if $(\Rsk , E, F)$ is achievable for some $E > 0$ and $F > 0$.
\end{definition}

%
We conclude our suite of definitions by formalizing the notion of degraded channels. 
\begin{definition}[Degradedness] \label{def:degraded_eve}
We say that the DMBC $p(x,y,z|s)$ is  \emph{degraded}  if  $(\xa,\xin) - \xb - \xe$ form a Markov chain, i.e., $p(y,z|x,s) = p(y|x,s)p(z|y)$.
\end{definition}
In this case, we also say that the DMBC $p(x,y,z|s)$ is degraded {\em in favor of} Bob or equivalently that Eve's observation is a {\em degraded version} of Bob's.  Note that we do not differentiate between  physical  and stochastic degradedness~\cite[Ch.\ 5]{ElGamal_Kim_LNIT}. The capacity results will turn out to be identical for both cases.

\section{Basic Capacity Results}
\label{sec:capacity_result}

We present our capacity results in this section. These correspond to Definitions~\ref{def:wach} and~\ref{def:cap} and we emphasize that $R_M$ and $R_\Phi$ are unconstrained here.   We leverage on a source emulation result  by Ahlswede-Csisz\'ar~\cite{Ahlswede_Csiszar93} to give a single-letter expression for the secret key capacity containing two auxiliary random variables taking into account that $S^n$ has to satisfy the cost constraint in~\eqref{equ:cost_constraint}. We also provide a looser  upper bound that contains no auxiliary random variables. The upper bound is tight when the DMBC  is degraded in favor of Bob. The capacity results in this section motivate the more refined error exponent analysis in the following section where $R_\Phi$ can be constrained and we will see that a judicious choice of $R_M$ does not reduce $C_{\mathrm{SK}}$ in the case of degraded DMBCs.

\begin{proposition}[Secret Key Capacity] \label{thm:capacity}
    The secret key capacity of DMBC $(\calS,p(x,y,z|s), \calX\times\calY\times\calZ)$ is
    \begin{equation} \label{equ:skey_capacity}
        \Csk (\Gamma) = \max  \,\,\, [I(U ;\xb|W) - I(U ;\xe|W) ] \ ,
    \end{equation}
    where the maximization is over  all joint distributions  that factor in accordance to $W-U-(X,S)-(Y,Z)$ or  equivalently,
\begin{align}
p(w,u,s,x,y,z) = p(w) p(u|w) p(x,s|u) p(y,z|x,s) \label{eqn:joint_dist}
\end{align}
such that $\rvE[\fcost(\xin)] \leq \Gamma$.
\end{proposition}

By repeated applications of Bayes rule, the decomposition in \eqref{eqn:joint_dist} can be written as
\begin{align}
p(w,u,s,x,y,z) = p(w|u) p(u|x,s) p(s) p(x,y,z|s) \ . \label{eqn:jd2}
\end{align}
Since the DMBC $p(x,y,z|s)$ is given, the optimization in~\eqref{equ:skey_capacity} is over the source distribution $p(s)$ and the auxiliary conditional distributions $p(w|u)$ and $p(u|x,s)$.   Furthermore, by using the Fenchel-Eggleston-Carath\'eodry Theorem~\cite[App.~C]{ElGamal_Kim_LNIT}, it can be argued that the cardinalities of the auxiliary random variables $W$ and $U$ can be bounded as $|\calW| \le |\calX||\calS|+3$ and $|\calU|\le (|\calX||\calS|+3)(|\calX||\calS|+1)$ respectively.

\begin{proof}[Proof of Proposition~\ref{thm:capacity}]
Achievability follows from \cite[Theorem~1]{Ahlswede_Csiszar93} for
Model SW with a slight modification to account for cost constraint on
$S^n$ in~\eqref{equ:cost_constraint}.  Fix an $\epsilon>0$ and a joint
distribution in \eqref{eqn:joint_dist} achieving $\rvE[\Lambda(S)]\le
\frac{\Gamma}{1+ \epsilon}$. Let $S\sim p_S(s)$ be the
$\calS$-marginal of \eqref{eqn:joint_dist} and let its typical
set\footnote{The {\em typical set} defined in
  $\calT_\epsilon^{(n)}(S)$~\cite[Sec.~2.4]{ElGamal_Kim_LNIT} consists
  of all sequences $s^n$ whose type (empirical distribution)
  $\pi(s;s^n)$ satisfies $|\pi(s;s^n)-p_S(s)|\le\epsilon\, p_S(s)$ for
  every $s\in\calS$. The typical average
  lemma~\cite[Sec.~2.4]{ElGamal_Kim_LNIT} implies that $n(1-\epsilon)
  H(S)\le \log|\calT_\epsilon^{(n)}(S)|\le n(1+\epsilon) H(S)$.}  be
$\calT_\epsilon^{(n)}(S)$. Index all the elements in
$\calT_\epsilon^{(n)}(S)$ as $[1: |\calT_\epsilon^{(n)}(S)| ]$. We are
only going to excite the DMBC $p(x,y,z|s)$ using sequences belonging
to $\calT_\epsilon^{(n)}(S)$. By the typical average
lemma~\cite[Sec.~2.4]{ElGamal_Kim_LNIT}, this ensures that for every
$n$, the almost sure cost constraint in~\eqref{equ:cost_constraint} is
satisfied.

The encoder has the codebook $\calT_\epsilon^{(n)}(S)$, which is known to all parties. Alice generates an index $M \in [1: |\calT_\epsilon^{(n)}(S)|]$    uniformly at random so in this coding scheme, $R_M  = \frac{1}{n}\log | \calT_\epsilon^{(n)}(S) |= H(S) + \delta(\epsilon)$ for some $\delta(\epsilon)\downarrow 0$ as $\epsilon\downarrow 0$.  Given  $M$, the encoder  transmits the sequence  indexed by $M$  in the codebook. Note that $p_S^n(\calT_\epsilon^{(n)}(S))$ is arbitrarily close to one for large enough $n$. Hence, just as in the proof of \cite[Theorem 1]{Ahlswede_Csiszar93}, we can consecutively select mutually disjoint wiretap codes  $\{\calC_i\}_{ i =1}^{ N}$ from $\calT_\epsilon^{(n)}(S)\times\calX^n$  (with $\eta$ in   \cite[Eq.~(4.1)]{Ahlswede_Csiszar93} replaced by $2\eta$, say) where each codebook $\calC_i$ contains codewords of  the same type. The rest of the proof  in~\cite[Theorem 1]{Ahlswede_Csiszar93}       follows verbatim with our $(X,S)$ in the role of $X$ there.
   This allows us to assert   that $I(U;Y|W)-I(U;Z|W)$ is a one-way (forward) achievable key rate. Note that in our setting,  Alice receives $X^n$ and also  has $S^n$ (a function of her   privately generated index $M$), Bob receives $Y^n$ and Eve receives $Z^n$.
The proof is completed by taking $\epsilon\downarrow 0$ and using the continuity of $\Gamma\mapsto C_{\mathrm{SK}}(\Gamma)$. That $C_{\mathrm{SK}}(\Gamma)$ is continuous follows from the continuity of $I(U;Y|W)$, $I(U;Z|W)$  and $\rvE[\Lambda(S)]$ in~\eqref{eqn:joint_dist}.

The converse proof of Theorem~\ref{thm:capacity} is  standard and we provide it in Section~\ref{sec:conv1} for completeness. It   relies on a simple application of the Csisz\'{a}r-sum-identity~\cite[Sec.\ 2.3]{ElGamal_Kim_LNIT} and an appropriate identification of the auxiliary random variables that satisfy the Markov conditions in \eqref{eqn:joint_dist}.
\end{proof}


To find the secret key capacity for specific channels,  two auxiliary random variables $W$ and $U$ solving~\eqref{equ:skey_capacity}    have to be identified. This may be a difficult task. In the next proposition, we provide an (albeit looser)  upper bound which does not involve any auxiliary random variables. This result will turn out to be important in Section~\ref{sec:example} where we present several channels for which we can calculate the secret key capacity-cost function in closed-form.

\begin{proposition}[Upper Bound in Secret Key Capacity]\label{thm:upper_bound}
    The secret key capacity is upper bounded as
    \begin{equation}\label{equ:upper_bound}
        \Csk(\Gamma) \leq \max \,\, I(\xa,\xin; \xb|\xe) \ ,
    \end{equation}
    where the maximization is over all input distributions $p(s)$ such that  $\rvE[\fcost(\xin)] \leq \Gamma$.
\end{proposition}
The proof of this proposition is given in Section~\ref{sec:loose_ub}. Roughly speaking, the expression in \eqref{equ:upper_bound} can be interpreted as the secret key capacity when Alice and Bob have full knowledge (side information) of Eve's observation $Z$, hence the conditioning on $Z$.  We note by using the techniques in Ahlswede-Csisz\'ar~\cite{Ahlswede_Csiszar93} (and in particular Lemma~2.2 therein) that our upper bound also holds for the scenario where the parties Alice and Bob can exchange {\em multiple} messages--the multi-way discussion scenario. 

In the case of degraded $p(x,y,z|s)$, the result in Proposition~\ref{thm:upper_bound} is tight.
\begin{corollary}[Secret Key Capacity of Degraded DMBCs] \label{corol:degraded_eve}
    If the DMBC  $p(x,y,z|s)$ is  \emph{degraded}, the secret key capacity is
    \begin{equation}\label{equ:degraded_capacity}
        \Csk(\Gamma) = \max \,\, \, [ I(\xa,\xin; \xb)-I(\xa,\xin;\xe)] \ ,
    \end{equation}
     where the maximization is over all input distributions $p(s)$ such that  $\rvE[\fcost(\xin)] \leq \Gamma$.
\end{corollary}

\begin{proof}
For achievability, we can choose $W=\varnothing$ and $U=(X,S)$ in~\eqref{equ:skey_capacity}. The Markov condition in~\eqref{eqn:joint_dist} is satisfied.

For the converse,   we observe from Proposition~\ref{thm:upper_bound} that  the secret key capacity of the degraded DMBC  can be upper bounded as
    \begin{align}
        \Csk(\Gamma) &\leq I(\xa,\xin; \xb | \xe)  \\
            &= I(\xa,\xin; \xb) - I(\xa,\xin;\xe)    \label{equ:degraded_upper_bound} \ .
    \end{align}
    The last equality is due to the fact that for degraded channels, $(\xa,\xin) - \xb - \xe$ forms a Markov chain.
\end{proof}

Notice that for a fixed $p(s)$,  the difference of mutual informations in \eqref{equ:degraded_capacity} can be decomposed into two parts:
\begin{align} \label{eqn:rate-separate}
I(X,S;Y)- I(X,S;Z) =R_{\mathrm{ch}}[p(s)] +  R_{\mathrm{src}}[p(s)],
\end{align}
where the channel and source rates are respectively defined as
\begin{align}
R_{\mathrm{ch}}[p(s)] &\triangleq  I(S;Y)-I(S;Z), \quad\mbox{and} \label{eqn:Rch}\\
 R_{\mathrm{src}}[p(s)] &\triangleq I(X;Y|S)-I(X;Z|S). \label{eqn:Rsrc}
\end{align}
The first rate $R_{\mathrm{ch}}[p(s)]$ can be interpreted as the confidential
message rate of the wiretap channel $p(y, z|s)$ \cite{Wyner75}. The second rate $R_{\mathrm{src}}[p(s)]$ is the secret key rate from an excited
correlated source $(X,  Y,Z)$ previously studied in \cite{chou_it10} for a
particular sounding signal $s^n$ with type $p(s)$. In the present setup, $s^n$ is randomly chosen by Alice. As such, we can optimize over its distribution
$p(s)$ to find the largest ``sum rate'' $R_{\mathrm{ch}}[p(s)] +  R_{\mathrm{src}}[p(s)] $. It turns out
that there is a natural interplay and tradeoff between $R_{\mathrm{ch}}[p(s)]$ and  $R_{\mathrm{src}}[p(s)]$. We illustrate
this numerically using an example in Section \ref{sec:binary}.

We provide an  alternative proof of the capacity of degraded DMBCs via the error exponent route in the next section. We note that the flexibility of the amount of private randomness that Alice has  in the form of the random message $M$ (which we did not exploit in this section) allows us to operate at a lower $R_{\Phi}$ and yet result in a positive capacity.

\section{Error Exponent Theorem}
\label{sec:exponent_result}

In this section, we present an inner bound for the secret key
capacity-reliability-secrecy region per Definition
\ref{def:re_region}. Our general result is then specialized to other
known results in the literature. Recall that for the error exponent
results, we consider the case when there is no cost constraint on the
codewords for simplicity (i.e., $\Gamma=\infty$). 



We make the following two observations when we employ the
achievability strategy proposed in this paper which is a random
binning scheme.  First, the decoding error probability
$\rvP(\KA\ne\KB)$ is only a function of marginal distribution
$p(x,y,s)=p(s)p(x,y|s)$.  Second, the key leakage $I( \KA;
\xe^n,\Phi)$ is only a function of marginal distribution $p(x,z,s)$.
This means that we can characterize the achievable reliability and
secrecy exponents separately as functions of each marginal
distribution.



\subsection{Basic Definitions}
Before we present our result, we begin with a few definitions.  Let
\begin{align}
& \tilEo^{(1)}( p(s),\rho, \Rpub )  \triangleq \nnum\\
&\quad\rho \Rpub   - \log \sum_{s,y} p(s)p(y|s) \left[\sum_{s,x}    p(x|y,s)^{\frac{1}{1+\rho}}\right]^{1+\rho} \hspace{-1em} , \label{eqn:tilEo1} \\
& \tilEo^{(2)}( p(s),\rho, \Rpub , R_M)  \triangleq \nnum\\
&\quad\rho (\Rpub -R_M)  - \log \sum_{s }   \left[\sum_{x,y}  p(s)  p(x,y|s)^{\frac{1}{1+\rho}}\right]^{1+\rho} \hspace{-1em} , \label{eqn:tilEo2} \\
& \tilEo^{(3)}( p(s),\rho, \Rpub, R_M)  \triangleq \nnum\\
&\quad\rho(\Rpub - R_M)  - \log \sum_y                         \left[\sum_{s,x} p(s)  p(x,y|s)^{\frac{1}{1+\rho}}\right]^{1+\rho} \hspace{-1em} . \label{eqn:tilEo}
\end{align}
As well, define
\begin{align}
&\Eo(p(s), \Rpub, R_M) \triangleq \min \Big\{  \max_{0\le\rho\le 1}  \tilEo^{(1)} (p(s),\rho, \Rpub ) , \nnum\\
&\,\,\max_{0\le\rho\le 1}  \tilEo^{(2)} (p(s),\rho, \Rpub, R_M ), \max_{0\le\rho\le 1}  \tilEo^{(3)} (p(s),\rho, \Rpub, R_M ) \Big\}.
   \label{eqn:relaible_expo} 
\end{align}
Similarly, define
\begin{align}
&\tilFo( p(s), \alpha, \Rsk, \Rpub, R_M) \triangleq   \nnum\\
&\!-\!\alpha(\Rsk\!+\! \Rpub \!-\! R_M)
       \!-\! \log \sum_{x,z,s}\! p(x,z,s)\!
                \left[\frac{p(x,z|s)}{p(z)}   \right]^{\alpha}   \!
        \label{eqn:tilFo}, \\
&\Fo(p(s), \Rsk, \Rpub, R_M) \!\triangleq \! \sup_{0<\alpha\le 1} \!\tilFo(  p(s),\alpha, \Rsk, \Rpub,R_M )    .   \label{eqn:secrecy_expo}
\end{align}
We now define a   rate-exponent region parameterized by the input distribution $p(s)$ and the pair of auxiliary rates $(\Rpub,R_M)$:
\begin{align}
    & \tilde{\calR}(p(s), \Rpub, R_M) = \Big\{ (\Rsk, \tilE , \tilF) \in \bbR_+^3:  \nnum \\
    & \qquad \tilE \leq \Eo(p(s),\Rpub, R_M)   \nonumber\\
    & \qquad \tilF \leq \Fo(p(s),   \Rsk, \Rpub, R_M) \Big\} \label{eqn:R_ps} \ .
\end{align}

\subsection{The Inner Bound}  \label{sec:inner}
The following theorem provides an inner bound to  the   capacity-reliability-secrecy region  $\calR$.

\begin{theorem}[Inner Bound to the Capacity-Reliability-Secrecy Region] \label{thm:inner_bound_triple}
The union of the regions in \eqref{eqn:R_ps} is an inner bound to the  secret key capacity-reliability-secrecy region, i.e.,
    \begin{equation}\label{eqn:inner_bound_triple}
        \bigcup_{p(s)  , \Rpub  , R_M  } \tilde{\calR}(p(s), \Rpub, R_M) \subseteq \calR \ .
    \end{equation}
\end{theorem}
The proof of this theorem can be found in Section \ref{sec:proof_exponent} and   hinges on an ML-MAP decoding strategy. More precisely, given $(y^n,\phi)$,  Bob  first uses the following rule to estimate Alice's source of private randomness  $\hatm$   and    Alice's received sequence $\hatx^n$:
\begin{equation} \label{eqn:mlmap1}
(\hatm,\hatx^n) \triangleq \argmax_{(m,x^n): \phi(m,x^n)=\phi  } p(y^n|s^n(m) ) p (x^n| y^n,s^n(m) ) \ .
\end{equation}
The function $\phi(m,x^n)$ is a (random) binning function, which is
defined and discussed in greater detail in Section~\ref{sec:defs}. The
exponents $\tilEo^{(1)}$ and $\tilEo^{(2)}$ represent the marginal
events $\{\hatM = M, \hatX^n\ne X^n\}$ and $\{\hatM \ne M, \hatX^n =
X^n\}$, respectively. The former is a Slepian-Wolf-type
exponent~\cite{gallager76} ($X$ to be reconstructed given vector
side-information $(Y,S)$) while the latter is a channel coding-type
exponent~\cite[Sec.~5.6]{gallagerIT} (input $S$ and vector output
$(X,Y)$). The exponent $\tilEo^{(3)}$ represents the joint error event
$\hatM\ne M,\hatX^n\ne X^n$ and is a hybrid of Slepian-Wolf and
channel coding.  Upon the decoding of $(\hatm,\hatx^n)$, Bob declares
his key to be $\kB = k(\hatm,\hatx^n)$, where $k(\cdot,\cdot)$ is another
(random) binning function.  The proof for the secrecy exponent
leverages on the properties of the R\'{e}nyi entropy as
in~\cite{Hayashi, chou_it10}.

The union of the regions in~\eqref{eqn:inner_bound_triple} is likely
to be a strict inner bound since our coding scheme does not involve
the use of any auxiliary random variables (unlike in
Proposition~\ref{thm:capacity}). However, as we shall see in
Section~\ref{sec:degr}, our analysis of the ML-MAP strategy shows that
all weakly-achievable rates $\Rsk < \Csk$ are strongly-achievable for
degraded channels. 





Another reason as to why the error exponent region is likely not tight
may be distilled from works by Csisz\'{a}r-Narayan
\cite{CsiszarN04}, later extended by Gohari-Anantharam~\cite{gohari_I, gohari_II}.  Consider an external agent who
can recover $X^n$ perfectly after receiving Eve's information
$(Z^n,\Phi)$ and the shared secret key $\KA$.  If the agent were not
able to recover $X^n$ there would be some piece of information about
$X^n$, independent of $(Z^n,\Phi,\KA)$, that the external agent would
require to know $X^n$ perfectly. In such a setting, Alice could reveal
the needed information on the public channel without lowering the
secret key rate.  This follows since what would be revealed is
independent of $\KA$, and thus of no use to Eve. Thus, without loss of
generality, we can assume the external agent knows $X^n$ perfectly.

Now, say that $Z$ is a degraded version of $Y$.  In this setting Bob
can simulate $Z^n$. Bob also has $(\Phi,\KB)$ (note that $\KB=\KA$
with high probability).  So, Bob too can be assumed to recover $X^n$
perfectly.  In other words, in the degraded setting there is no loss
in generality in requiring Bob to recover $X^n$. However, when there
is a non-trivial joint distribution amongst $X,Y$ and $Z$ (i.e., the
non-degraded case), it is not necessarily true that Bob can recover
$X^n$. Hence the error-exponent strategy may be strictly suboptimal
(at least in a capacity sense for non-degraded channels). This
observation is consistent with the ``separation'' strategy elucidated
in \eqref{eqn:Rch} and \eqref{eqn:Rsrc} as the separation
strategy--which is optimal in the degraded case--in effect implies
that Bob can decode $X^n$ as discussed in the previous paragraph.


\subsection{Positivity of Error Exponents and Interpretations}
\begin{figure}[t]
  \centering
    \begin{overpic}[width=.99\columnwidth]{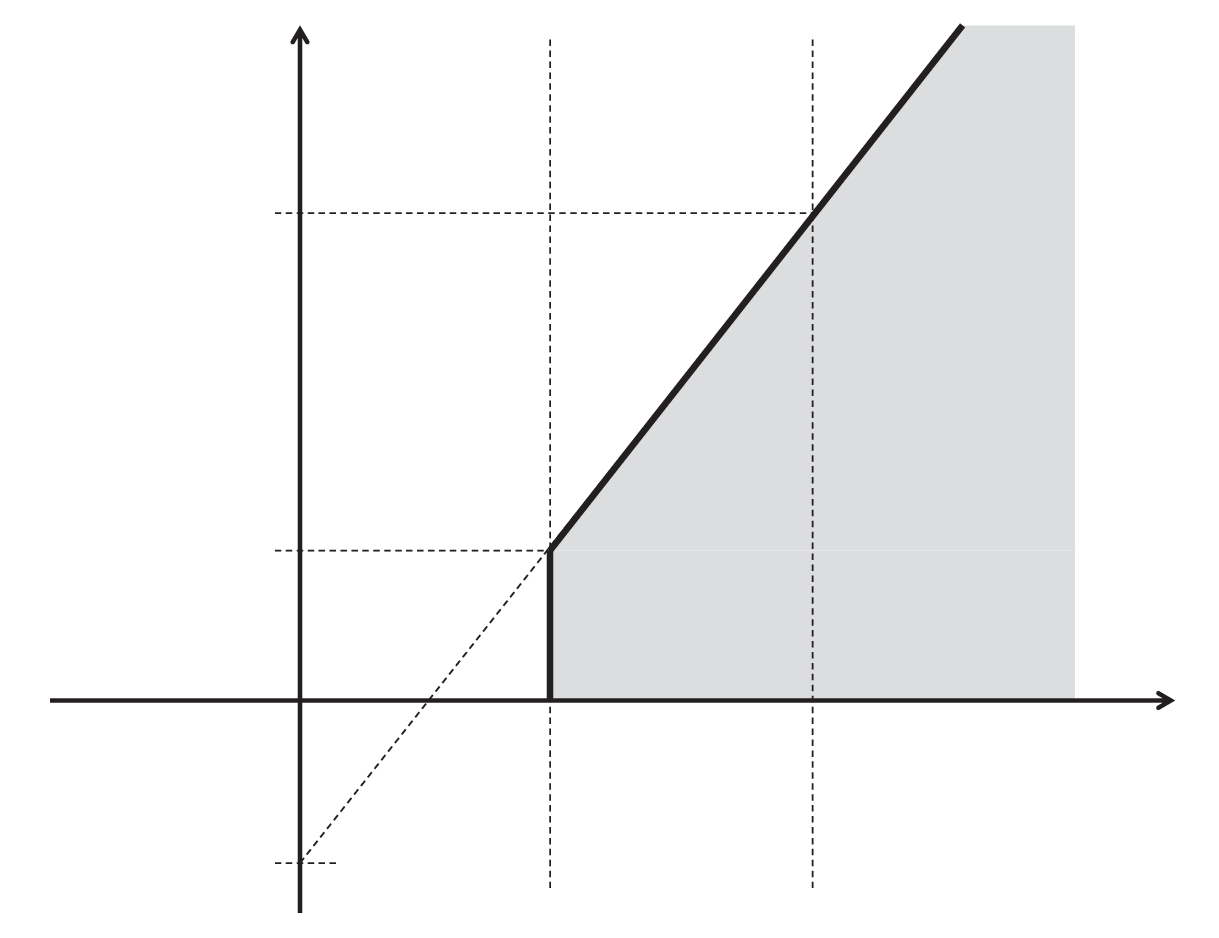}
    \put(92,21){\small $R_\Phi$}
    \put(17,72){\small $R_M$}
    \put(13,59){\small $H(S)$}
    \put(10,31){\small $I(S;Y)$}
    \put(10,5){\small $I(S;Y)$}
    \put(20,15){\small $0$}
    \put(2,1){\small $ -H(X|Y,S)$}
    \put(30,0){\small $H(X|Y,S)$}
    \put(60,0){\small $H(X,S|Y)$}
    \put(36,28){\rotatebox{52}{\small  $R_M = R_\Phi+ I(S;Y) -H(X|Y,S)$}}
     \put(55,30){\large ${\calE}_{\mathrm{o}}^+$}
    \put(47,28){$\large \rvA$} 
    \put(45,31){\circle*{2}}
    \put(67,59){\circle*{2}}
    \put(68,56){$\large \rvB$} 
    \end{overpic}
      \caption{The region where  $\Eo(p(s),\Rpub, R_M)$ is positive is denoted by the shaded set ${\calE}_{\mathrm{o}}^+$. See \eqref{eqn:pos_reliability_phi} and \eqref{eqn:pos_reliability}. Points $\rvA=(H(X|Y,S),I(S;Y))$ and $\rvB=(H(X,S|Y), H(S))$ respectively denote the two-step approach (of Bob first recovering $M$ through channel decoding and then recovering $X^n$ via Slepian-Wolf decoding) and the source emulation approach (with vector source $(X,S)$ given $Y$ just as in the achievability  proof of Proposition~\ref{thm:capacity}) discussed in greater detail in Section~\ref{sec:connect_to_previous}-III. The semi-infinite ray emanating from $\rvA$, passing through $\rvB$, and extending northeast is the capacity-achieving set of $(R_\Phi, R_M)$ for our error exponent scheme. For source emulation it  only starts  from $\rvB$ and extends northeast.}
\label{fig:pos_E0}
\end{figure}
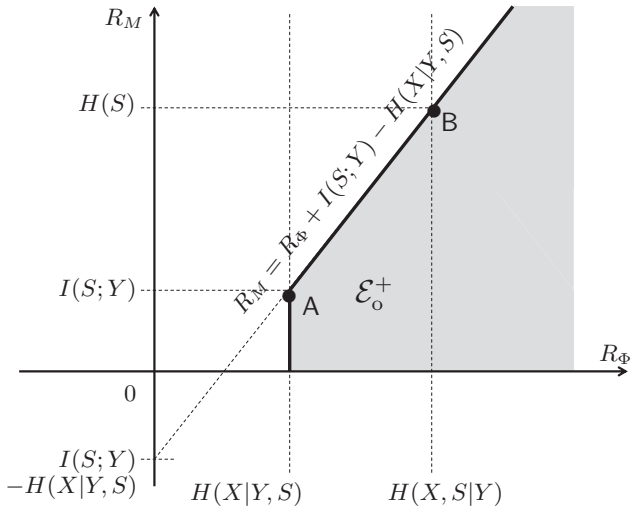

\begin{figure}[t]
  \centering
    \begin{overpic}[width=.99\columnwidth]{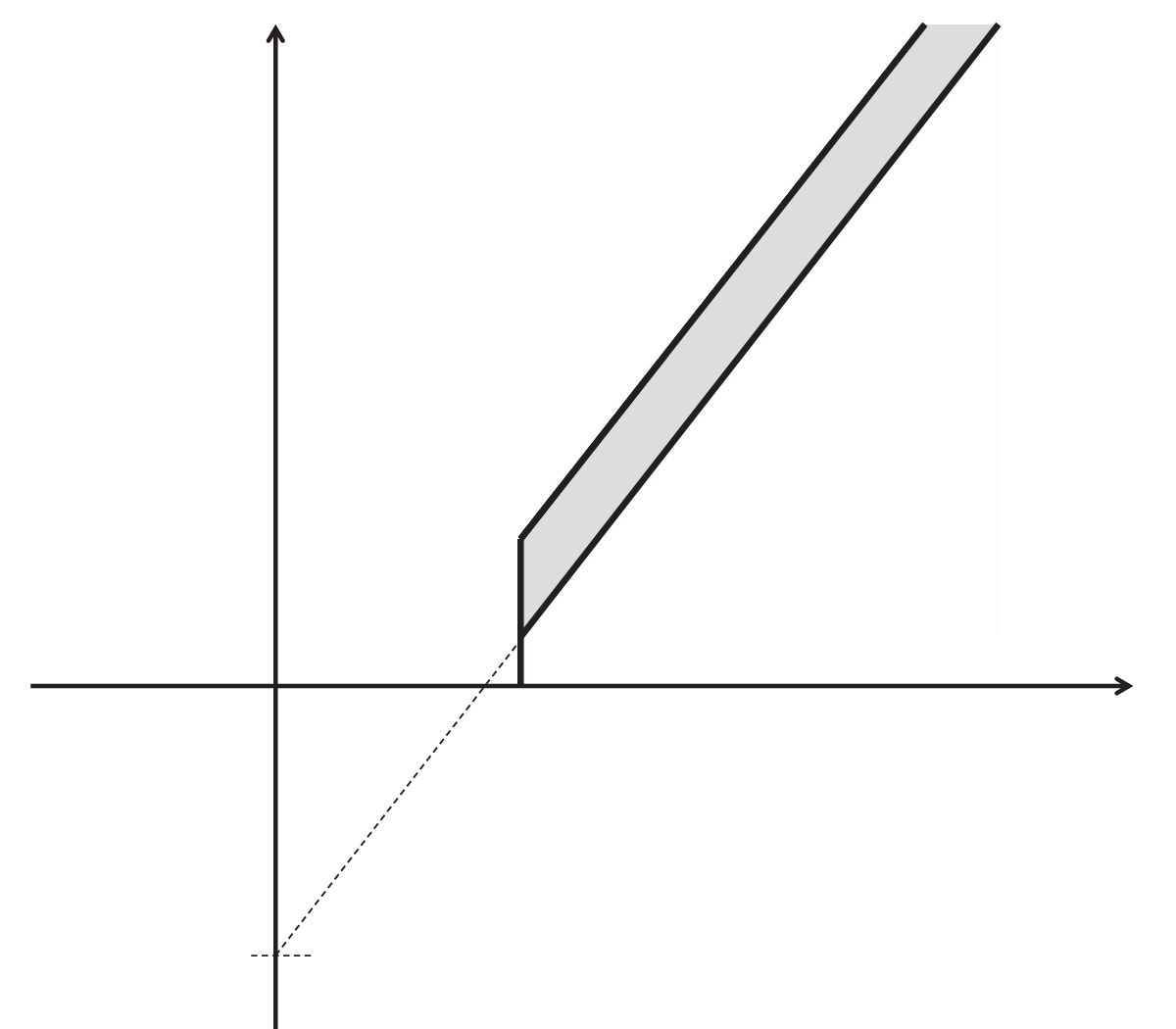}
    \put(93,32){\small $R_\Phi$}
    \put(15,84){\small $R_M$}
    \put(20,25){\small $0$}
    \put(46,32){\rotatebox{52}{\small  $R_M = R_\Phi +\Rsk + I(S;Z) -H(X|Z,S)$}}
    \put(0,7){\small $\Rsk\! +\! I(S;Z)$}
    \put(0,2){\small $ -H(X|Z,S)$}
    \put(45,70){\large ${\calE}_{\mathrm{o}}^+\cap{\calF}_{\mathrm{o}}^+$}
    \put(52,68){\vector(1,-1){9}}     
    \end{overpic}
      \caption{This is the same as Fig.~\ref{fig:pos_E0} with \eqref{eqn:pos_secrecy} also illustrated. The region where  $\Eo(p(s),\Rpub, R_M)$ and  $\Fo(p(s),\Rpub, R_M)$ are both positive is denoted by the shaded set ${\calE}_{\mathrm{o}}^+\cap{\calF}_{\mathrm{o}}^+$. This combines the rate constraints in \eqref{eqn:pos_reliability_phi}, \eqref{eqn:pos_reliability} and \eqref{eqn:pos_secrecy}.  The intuition here is the following: To maximize  $\Rsk$, the line indicated by the equation  $R_M = R_\Phi +\Rsk + I(S;Z) -H(X|Z,S)$ should be shifted upwards until the shaded region almost vanishes.}
\label{fig:pos_E0_F0}
\end{figure}
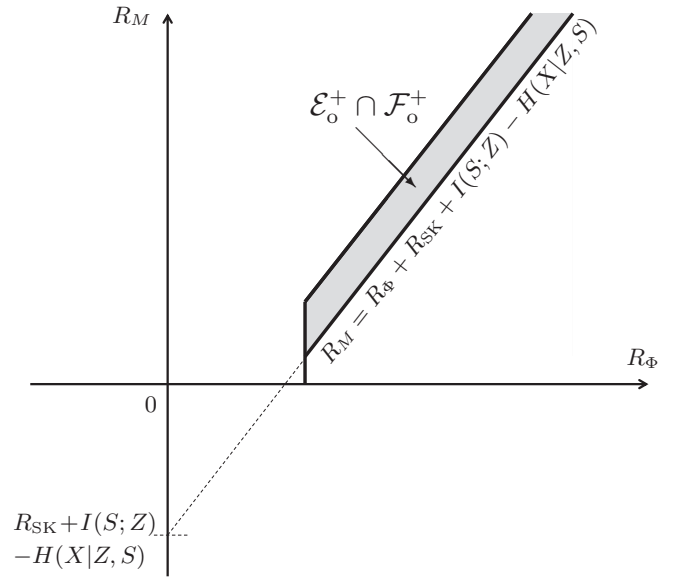

For a particular choice of input distribution $p(s)$, the following proposition characterizes the boundary of the achievable rate-exponent region  in~\eqref{eqn:R_ps}.
\begin{proposition}[Positivity of  Error Exponents]\label{prop:inner_bound_boundary}
    For a fixed    $p(s)$, the exponent $\Eo(p(s),\Rpub, R_M)$ in~\eqref{eqn:relaible_expo} is positive if
    \begin{align}
                R_{\Phi}  &  > H(X|Y,S)\quad     \mbox{and} \label{eqn:pos_reliability_phi}\\
                R_{\Phi} - R_M   &> H(X|Y,S)  - I(S;Y)  \ . \label{eqn:pos_reliability}
    \end{align}
See Fig.~\ref{fig:pos_E0}.    Similarly,  the exponent $\Fo(p(s),  \Rsk, \Rpub, R_M)$ in~\eqref{eqn:secrecy_expo} is positive if
    \begin{equation}
        \Rsk+ R_{\Phi} - R_M  < H(X|Z,S ) - I(S;Z) \ . \label{eqn:pos_secrecy}
    \end{equation}
See Fig.~\ref{fig:pos_E0_F0}.
\end{proposition}


The proposition can be proved by firstly verifying that $\tilEo^{(j)},
j=1,2,3$ (resp.\ $\tilFo$) are concave functions of $\rho$
(resp.\ $\alpha$); secondly by computing the partial derivative of
$\tilEo^{(j)}$ (resp.\ $\tilFo$) with respect to $\rho$
(resp.\ $\alpha$); and finally by evaluating the slope at $\rho=0$
(resp.\ $\alpha=0$).  This is a standard calculation and as such, we
omit the details. See~\cite[Theorem~3]{chou_it10} and the accompanying
remarks for similar calculations. Note that there are only two rate
constraints for reliability in \eqref{eqn:pos_reliability_phi} and~\eqref{eqn:pos_reliability}. This is because the rate constraint  required for  $\tilEo^{(2)}>0$ is
\begin{equation}
R_M -R_\Phi   <  I(X,Y;S) \label{eqn:pos_E2}
\end{equation}
which is already implied by \eqref{eqn:pos_reliability} since $I(S;Y)
- H(X|Y,S) = I(X,Y;S) - H(X|Y) \le I(X,Y;S)$.  Note that in the
derivation of $\tilEo^{(2)}$ and \eqref{eqn:pos_E2}, we treat $(X,Y)$
as a vector output of a channel with input $S$.  We had mentioned
previously that $R_\Phi$ can be reduced and yet the secret-key
capacity would remain unchanged if we reduce $R_M$
accordingly. However, we observe from \eqref{eqn:pos_reliability_phi}
that there is nevertheless a lower bound on $R_\Phi$ due to a marginal
error event.  Thus, $R_\Phi$ cannot be reduced arbitrarily, and in
particular not beyond the conditional entropy $H(X|Y,S)$. Intuitively,
the corner point in Fig.~\ref{fig:pos_E0} (point $\rvA$) where $R_\Phi
= H(X|Y,S)$ and $R_M = I(S;Y)$ may be achieved  from a two-step decoding
procedure where Bob first recovers $M$ through channel decoding given
$Y^n$ and then recovers $X^n$ via Slepian-Wolf decoding given the
vector side-information $(S^n(M),Y^n)$ ($M$ assumed to be decoded
correctly). This two-step decoding procedure is, however, not what we do in the ML-MAP decoding scheme in  \eqref{eqn:mlmap1}. The ML-MAP decoding scheme decodes $M$ and $X^n$ {\em jointly} so its exponent is likely to be  higher than the two-step decoding scheme.



The first rate condition in~\eqref{eqn:pos_reliability} for the
reliability exponent to be positive may be rewritten as follows:
\begin{align}
R_M  < I(S;Y) + [R_{\Phi}- H(X|Y,S)]   \ . \label{eqn:RM_less} 
\end{align}
Using~\eqref{eqn:RM_less},  we see that if $R_{\Phi}>H(X|Y,S)$ (i.e., the
compression rate is strictly larger than the Slepian-Wolf limit
$H(X|Y,S)$  as allowed by
\eqref{eqn:pos_reliability_phi}), we may transmit the message $M$
reliably at rates higher than $I(S;Y)$, which is the maximum
transmission rate when the input distribution $p(s)$ is used for the
channel $p(y|s)$. 


The rate condition in~\eqref{eqn:pos_secrecy} for the secrecy exponent
to be positive may   be written in the following   equivalent
forms:
\begin{subequations}
\begin{eqnarray}
\hspace{-.2in}\Rsk+ R_{\Phi} \!\!\!\!&<& \!\!\!\! H(X|Z,S) + [R_M - I(S;Z) ] \ , \label{eqn:Rsk_less} \\
\hspace{-.2in} R_M  \!\!\!\!&> & \!\!\!\! I(S;Z) \!-\! [H(X|Z,S) \!-\! (\Rsk \! +\! R_{\Phi} ) ]   . \label{eqn:RM_more}
\end{eqnarray}
\end{subequations}
The authors in \cite[Theorem~3]{chou_it10} showed that the secrecy
exponent is positive when $\Rsk+ R_{\Phi}<H(X|Z,S)$. However, we
observe from \eqref{eqn:Rsk_less} that if $R_M>I(S;Z)$ (i.e., the
message rate is larger than what Eve can resolve with her channel
$p(z|s)$), the secrecy exponent is positive even though $\Rsk+
R_{\Phi}$ may be larger than $H(X|Z,S)$.  Similarly, observe
from~\eqref{eqn:RM_more} that if $\Rsk+ R_{\Phi}<H(X|Z,S)$, then $R_M$
may be smaller than $I(S;Z)$ for the secrecy exponent to be positive.


\subsection{Strong Achievability and Connections to Degradedness} \label{sec:degr}

Assume  that the DMBC $p(x,y,z|s)$ is degraded. We then eliminate the  rates $R_{\Phi}$ and $R_M$ in~\eqref{eqn:pos_reliability} and~\eqref{eqn:pos_secrecy} and  conclude  that $\Rsk$ is strongly-achievable   if
\begin{align}
    \Rsk &< H(X|Z,S ) - I(S;Z) - (H(X|Y,S ) - I(S;Y)) \nnum   \\
    & =  I(X;Y|S) - I(X;Z|S) - I(S;Z)  + I(S;Y)  \nnum  \\
    & =  I( X,S;Y) - I(X,S;Z)    \label{eqn:diff_mis} \nnum\\
        & =  I( X,S;Y|Z) \ ;
\end{align}
per~(\ref{eqn:pos_reliability_phi}) we also require that $R_\Phi >
H(X|Y,S)$.  The last equality holds due to the assumption of
degradedness, cf. Defn.~\ref{def:degraded_eve}. See Fig.~\ref{fig:pos_E0_F0}. This concurs with the
result for the secret key capacity for degraded channels obtained
using pure source emulation in Corollary~\ref{corol:degraded_eve}.
This alternative method of deriving the secret key capacity for the
degraded case via the error exponent route demonstrates that for
degraded channels, the weak and strong definitions for achievability
(in Definitions~\ref{def:wach} and~\ref{def:strong_achievable}
respectively) coincide.

\subsection{Connections to Previous Results}
\label{sec:connect_to_previous}



The reliability exponent in~\eqref{eqn:tilEo} is akin to a combination
of Gallager's exponents for channel
coding~\cite[Sec.\ 5.6]{gallagerIT} and for source coding with side
information~\cite{gallager76}. The secrecy exponent has been studied
for the secret key agreement source model~\cite{BBCM95, Hayashi}, the
corresponding channel model \cite{Hayashi}, and the source model with
external deterministic excitation~\cite{chou_it10}.
Hayashi~\cite{Hayashi, Hayashi06} also analyzed the exponential decay
of the information leakage rate for the wiretap channel.  The
expression in \eqref{eqn:tilFo} is akin to a combination of the key
leakage rate due to Eve's DMC $p(z|s)$~\cite{Hayashi} and the secrecy
exponent of the excited DMMS $p(x,z|s)$~\cite{chou_it10}.

In light of these observations,
Proposition~\ref{prop:inner_bound_boundary} may be specialized to
derive conditions for the positivity of the exponents for the pure
channel-type and the pure source-type models:

\begin{table}[!t]
\caption{Specialization of Proposition~\ref{prop:inner_bound_boundary} to existing results}
\label{table:special_EoFo}
\centering
\begin{tabular}{|c|c||c|c|} \hline
   & Specialization & Reliability $\Eo$ & Secrecy $\Fo$  \\
  \hline
  I & $\xa=\varnothing$     & Channel coding        & Wiretap channel  \\
  & $\Rpub=0$             & \cite[Theorem~5.6.2]{gallagerIT}     &  coding \cite[Theorem  3]{Hayashi}\\
  \hline
  II & $\xin=\varnothing$    & Source coding with                   &  Secret key generation with \\
     & $R_M=0$               & side information \cite{gallager76}   &  public discussion \cite{Hayashi} \\
  \hline
   &          & Source emulation              & Source emulation  \\
    III  & $R_M=H( S )$                    & $(X, S )$                       & $(X, S )$ \\
      &     & applied to \cite{gallager76}  & applied to \cite{Hayashi} \\
  \hline
\end{tabular}
\end{table}

\begin{enumerate}[I.]
  \item \emph{Alice has no access to the channel output ($\xa
    \leftarrow\varnothing$) and no public discussion ($\Rpub=0$)}:
    This case specializes to the wiretap channel $p(y,z|s)$. In this
    case, the reliability exponent $\Eo(p(s), 0, R_M)$ reduces to that
    of channel coding over a discrete memoryless channel (DMC)
    \cite[Theorem 5.6.2]{gallagerIT} and~\eqref{eqn:pos_reliability}
    reduces to the condition
      \begin{equation}
       R_M < I(S;Y),
      \end{equation}
      which we recognize as the condition for reliable communication
      over the DMC $p(y|s)$.

      In addition, our secrecy exponent $\Fo(p(s), \Rsk, 0, R_M)$
      reduces to Hayashi's wiretap secrecy exponent
      in~\cite[Eq.~(14)]{Hayashi} and~\eqref{eqn:diff_mis} reduces to
      the confidential message rate constraint
      \begin{equation}
      \Rsk < I(S;Y) - I(S;Z), \label{eqn:wiretap_special}
      \end{equation}
      which we recognize as the condition for reliable communication
      and secrecy for the wiretap channel. Note that the usual
      auxiliary random variable ``$U$'' \cite[Theorem
        22.1]{ElGamal_Kim_LNIT} has been taken to be equal to the
      source $S$ in~\eqref{eqn:wiretap_special}.
  \item \emph{Alice has no control of the channel input}: This case
    specializes to the secret key generation model with public
    discussion characterized by the DMMS $p(x,y,z)=\sum_s p(s)
    p(x,y,z|s)$ studied
    in~\cite{maurer93,CsiszarN04,CsiszarN08,gohari_I, gohari_II}.  The
    reliability exponent was characterized in~\cite{gallager76} and
    was stated as a special case of the main result
    in~\cite{chou_it10}. By letting $\xin\leftarrow\varnothing$ and
    $R_M=0$, \eqref{eqn:pos_reliability} simplifies to
      \begin{equation}
      R_{\Phi}    > H(X|Y) \label{eqn:sw}
      \end{equation}
      which we recognize as the condition for lossless source coding
      of $X$ given side information $Y$ \cite{SlepianWolf}.  This
      recovers an analogue of the result
      in~\cite[Theorem~3]{chou_it10}. Inequality \eqref{eqn:sw} also
      concurs with~\eqref{eqn:pos_reliability_phi}.

      We remark that Watanabe et al.~\cite{Wat10} showed that strongly
      secure privacy amplification is not achievable by Slepian-Wolf
      coding. But this does not contradict our error exponent result
      because the codes used in \cite{Wat10} have rates tending to the
      optimal compression rate $H(X|Y)$ in \eqref{eqn:sw} at a rate of
      $b/\sqrt{n}$ for some $b \in \bbR$ (cf.~\cite{TK12}). However,
      we operate at rates {\em strictly above} $H(X|Y)$
      in~\eqref{eqn:sw} so strong secrecy is indeed possible.

      The secrecy exponent $\Fo(p(s), \Rsk, R_{\Phi}, 0)$ was derived
      in~\cite{BBCM95, Hayashi, chou_it10}. Our secrecy exponent
      result in~\eqref{eqn:pos_secrecy} specializes in this case to
      \begin{equation}
         \Rsk+ R_{\Phi}    < H(X| Z )
      \end{equation}
      which recovers an analogue of the main result in Chou {\em et al.} \cite[Theorem~3]{chou_it10}.

  \item \emph{Alice excites the channel with $S^n$ generated in an
    i.i.d. manner according to $p_S$ and considers the joint variable
    $(X, S )$ as her source}: This is similar to the source emulation
    scheme adopted in the proof of Proposition \ref{thm:capacity}
    without cost constraint and ignoring the encoder but considering
    the three terminals: Alice with $(X, S )$, Bob with $Y$, and Eve
    with $Z$. This is point $\rvB$ in Fig.~\ref{fig:pos_E0}.  The
    reliability and secrecy exponents will be of the form in
    \cite{gallager76} and \cite{Hayashi}, respectively, with
    i.i.d.\ source $(X, S )$.  Thus substituting $R_M=H( S )$
    in~\eqref{eqn:pos_reliability} and \eqref{eqn:pos_secrecy} yields
      \begin{align}
        R_{\Phi}          &> H(X |Y, S )-I( S ;Y)+H( S ) \nnum\\
         &=H(X, S |Y) \label{eqn:src_emulation_relCond} \\
        \Rsk + R_{\Phi}   &< H(X |Z, S )-I( S ;Z)+H( S )\nnum\\
         &=H(X, S |Z)          \label{eqn:src_emulation_secCond} \ .
      \end{align}
      Upon  the elimination of $R_\Phi$  which, by \eqref{eqn:src_emulation_relCond}, satisfies the required lower bound  in \eqref{eqn:pos_reliability_phi}, we have 
      \begin{align}
      \Rsk &< H(X, S |Z) - H(X, S |Y)\nnum  \\
      & = I(X, S ;Y)-I(X, S ;Z) \ . \label{eqn:eliminate2}
      \end{align}
Notice that the difference of mutual informations on the RHS of
\eqref{eqn:eliminate2} is $I(X, S ;Y|Z)$ for degraded DMBCs. This
concurs with the secret key capacity of degraded DMBCs in
Corollary~\ref{corol:degraded_eve}.  

As is mentioned in the Introduction, while the source emulation scheme
achieves the secret key capacity, this rate cannot be strongly
achieved (per Definition~\ref{def:strong_achievable}) if $R_{\Phi}$ is
upper bounded by some quantity (but nonetheless still satisfies the
lower bound in~\eqref{eqn:pos_reliability_phi}) if we do not also have
the flexibility to concurrently set the rate of the sounding signal
$R_M$.  Observe that the lower bound on $R_\Phi$ in
\eqref{eqn:src_emulation_relCond} resulting from the pure source
emulation strategy (cf.~the achievability proof of
Proposition~\ref{thm:capacity}) is $H(X,S|Y)$ which is at least as
large as $H(X|Y,S)$ in \eqref{eqn:pos_reliability_phi} in
Proposition~\ref{prop:inner_bound_boundary} and, in general, is
strictly larger. Thus, our error exponent scheme which involves
wiretap coding plus key distillation allows us to reduce $R_\Phi$ from
$H(X,S|Y)$ to $H(X|Y,S)$--the difference being $H(S|Y)$.

\end{enumerate}
The specializations are summarized in Table \ref{table:special_EoFo}.

\section{Numerical Examples}
\label{sec:example}

We consider two examples in this section. The first  example illustrates the tradeoffs involved in the  capacity results in Section~\ref{sec:capacity_result}.  The second example illustrates the tradeoffs in the achievable error exponent results in Section~\ref{sec:exponent_result}.
\subsection{Capacity of the Binary On-off Channel} \label{sec:binary}
\begin{figure}
  \center
  \includegraphics[width=0.45\textwidth]{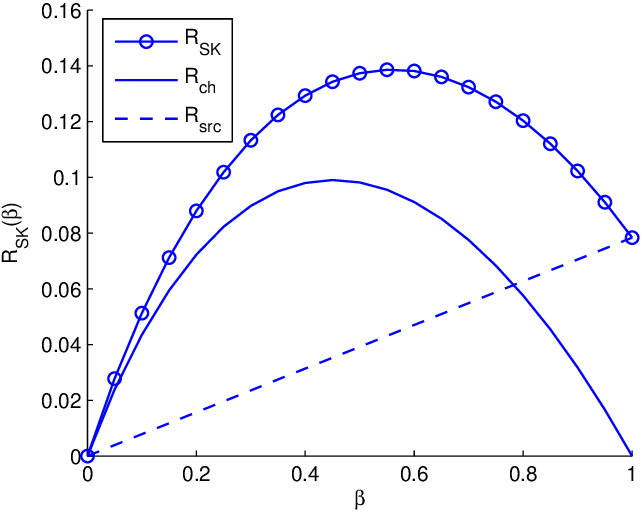}
  \caption{Secret key rate of  the binary on-off channel as a function of $\beta$. The input $S\sim \randBern{\beta}$. The parameters are $q=0.5, \tilq=0.8$, $\delta=0.1, \delta_3=0.2$. Note that $C_{\mathrm{SK}} = \max_{\beta \in [0,1]} R_{\mathrm{SK}}(\beta)$ and the maximizing $\beta^*\approx 0.59$.}
  \label{fig:binary_onoff}
\end{figure}

For our first example consider the binary on-off model
\begin{eqnarray*}
  \xa &=& H\cdot \xin \oplus N_1\\
  \xb &=& H\cdot \xin \oplus N_2\\
  \xe &=& ( \tilH \cdot H) \cdot  \xin \oplus N_3 \ ,
\end{eqnarray*}
where all the variables are binary and where the operations are
performed in the field of size 2. Hence, the addition above is is
binary modulo-2 addition. The ``channel gain'' $H$ is $\randBern{q}$
and $\tilH$ is $\randBern{\tilq}$.\footnote{We say that a binary
  random variable $X$ is $\randBern{\gamma}$ if $\Pr[X=1] = \gamma$.}
Noise $N_i$ is $\randBern{\delta_i}$ and the $N_i$ are mutually
independent. The channel describes a model in which, in the absence of
noise, Eve's observation is strictly worse than that of Alice's and
Bob's since $\tilH$ is present.

If $\delta_1=\delta_2=\delta$ and $\tilq \delta< \delta_3$, then Eve's
channel output is a degraded version of Bob's. In this case, there
exists a $Z' \triangleq \tilH' \cdot \xb \oplus N'_3$ for some
$\tilH'$, with the same distribution as $\tilH$, and independent $N'_3
\sim \randBern{\delta'_3}$ such that $(\xa, \xin) - Y - Z'$, where
\begin{equation}
    \delta'_3 = \frac{\delta_3 - \tilq \delta}{1 - 2 \tilq \delta}  \nonumber \ .
\end{equation}

Let $\xin\sim \randBern{\beta}$. The first term of $R_{\mathrm{ch}}$   is
\begin{align*}
    &I(\xin; \xb) = H(\xb) - H(\xb | \xin) \\
    &= \Hb(\beta q * \delta) - [\beta H(\xb|\xin=1) + (1-\beta) H(\xb|\xin=0)] \\
    &= \Hb(\beta q * \delta) - \beta \Hb(q * \delta) -(1-\beta)\Hb(\delta) \ ,
\end{align*}
where $\Hb(\cdot)$ is the binary entropy function and the operation $a*b \triangleq a(1-b)+(1-a)b$.
Similarly, the second term of $R_{\mathrm{ch}}$ can be expressed as
\begin{align*}
    I(\xin; \xe) = \Hb(\beta \tilq q * \delta_3) - \beta \Hb(\tilq q * \delta_3) -(1-\beta)\Hb(\delta_3) \ .
\end{align*}
The secret key rate due to source $\xa$ can be calculated as
\begin{align*}
    &R_{\mathrm{src}} = I(\xa; \xb|\xin) - I(\xa;\xe|\xin) \\
    &=\beta [I(\xa; \xb|\xin=1) - I(\xa;\xe|\xin=1) ] \\
    &= \beta [ \Hb(q*\delta) - \Hb(\delta * \delta) - \Hb(\tilq q * \delta_3) \\
    &\hspace{6ex}   + (1- q*\delta)\Hb(\delta'_3) + (q*\delta) \Hb(\tilq * \delta'_3)] \ .
\end{align*}
The second equality follows because if $S=0$, the source is not
observed and so there is no mutual information between $X$ and $Y$
(nor between $X$ and $Z$).

The secret key rate when the input is a $\randBern{\beta}$ source is
$\Rsk(\beta) = R_{\mathrm{ch}}(\beta) + R_{\mathrm{src}}(\beta)$ which
is plotted in Fig.~\ref{fig:binary_onoff} as a function of $\beta$
for the following parameters: $q=0.5, \tilq=0.8$, $\delta=0.1,
\delta_3=0.2$. Note that $R_{\mathrm{ch}}$ is a concave function of
$\beta$ while $R_{\mathrm{src}}$ is a linear function of $\beta$. If
$\beta=0$ then $\Rsk = 0$ since $X,Y,Z$ are jointly statistically
independent.  On the other hand, if $\beta=1$ then $\xin^n$ is the all
ones sequence and the $R_{\mathrm{src}}$ is maximal since the input
excites all common randomness due to the {\em common} on-off
coefficient $H$. However, when $\beta=1$, the secrecy rate of the
wiretap channel $R_{\mathrm{ch}}=0$. As we decrease $\beta$
$R_{\mathrm{ch}}$ initially increases faster than $R_{\mathrm{src}}$
decreases, resulting in the maximum $\Rsk$ being achieved at an
intermediate value of $\beta$.  In this example we have observed an
inherent tradeoff between the amount of the secret key rate due to
common randomness and due to wiretap secrecy.


\subsection{Error Exponents} \label{sec:num_ee}

\begin{figure}
  \centering
  \includegraphics[width=0.49\textwidth]{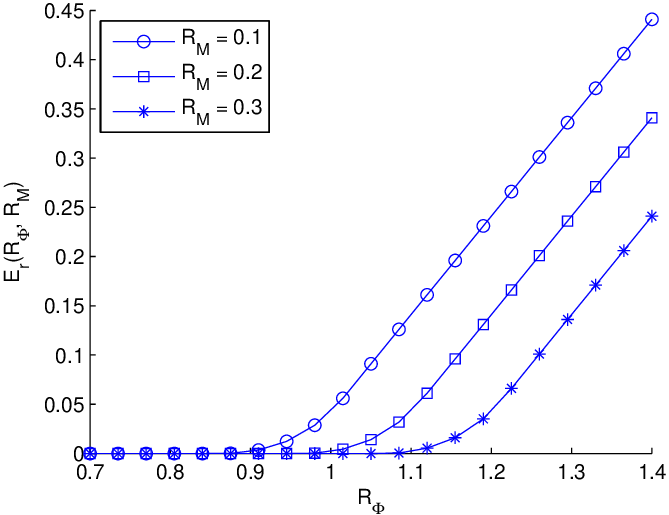}
  \caption{Plot of the random coding reliability exponent $\Er$ in \eqref{eqn:Er} }
  \label{fig:Erplot}
\end{figure}

We now illustrate our error exponent results. We assume that all
variables are binary valued, i.e.,
$\calX=\calY=\calZ=\calS=\{0,1\}$. We selected the parameters of the
DMBC $p(x,y,z|s)$ to ensure that Eve's observation $Z$ is a degraded
version of Bob's $Y$. We do so by first selecting the parameters of
the conditional distribution $p(x,y|s)$, then we proceeded to choose
the parameters in the conditional distribution $p(z|y)$. We keep the
channel $p(x,y,z|s)$ fixed throughout this subsection. Define the {\em
  input distribution-optimized reliability exponent}
\begin{equation}
\Er(R_{\Phi},R_M) \triangleq \max_{p(s)} \Eo(p(s), R_{\Phi},R_M)   \ ,\label{eqn:Er}
\end{equation}
where $\Eo$ was defined in \eqref{eqn:relaible_expo}. Also define the {\em  input distribution-optimized secrecy exponent}:
\begin{equation}
\Fr(\Rsk, R_{\Phi},R_M) \triangleq \max_{p(s)} \Fo(p(s),\Rsk ,  R_{\Phi},R_M)   \ , \label{eqn:Fr}
\end{equation}
where $\Fo$ was defined in \eqref{eqn:secrecy_expo}.  Note that for a particular set of rates $(\Rsk, R_{\Phi}, R_M)$, the optimal input distributions $p^*(s)$ in~\eqref{eqn:Er} and~\eqref{eqn:Fr} may be {\em different}. Hence, one has to use a {\em common} $p(s)$ in \eqref{eqn:inner_bound_triple}. We append the subscript $\mathrm{r}$ to $\Er(R_{\Phi},R_M)$ and $\Fr(\Rsk, R_{\Phi},R_M)$ to allude to the fact that in the derivation of these exponents, we use both {\em random coding}~\cite{gallagerIT} and {\em random binning} schemes~\cite{gallager76}.

\begin{figure}
  \centering
  \includegraphics[width=0.5\textwidth]{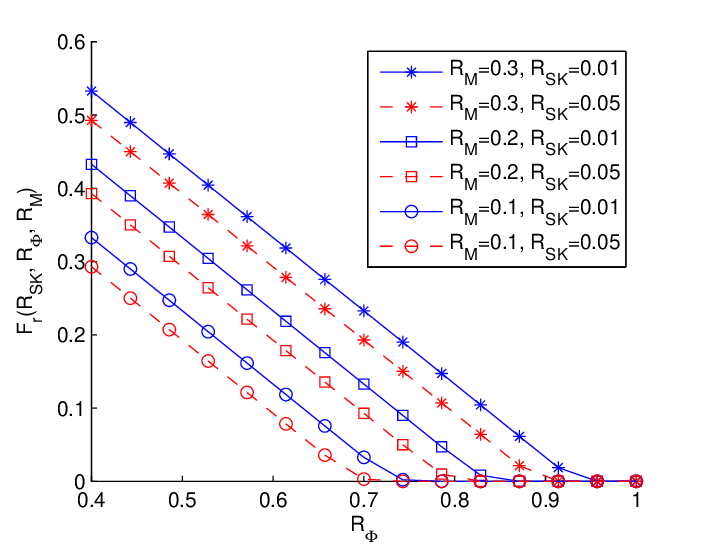}
  \caption{Plot of the random coding secrecy exponent $\Fr$ in \eqref{eqn:Fr} }
  \label{fig:Frplot}
\end{figure}

The functions $\Er(R_{\Phi},R_M)$ and $\Fr(\Rsk, R_{\Phi},R_M)$ are
plotted in Figs.~\ref{fig:Erplot} and~\ref{fig:Frplot}
respectively. From Fig.~\ref{fig:Erplot}, we observe that
$R_{\Phi}\mapsto \Er(R_{\Phi},R_M)$ is a non-decreasing function. This
is intuitive because given more information (i.e., when $R_{\Phi}$ is
large) and with $R_M$ fixed, Bob can decode the key $\KB$ with greater
reliability. In contrast, $R_M\mapsto \Er(R_{\Phi},R_M)$ is a
non-increasing function. This is also intuitive because Alice's
private source of randomness is increased if $R_M$ is increased making
it more challenging for Bob to decode the key.


From Fig.~\ref{fig:Frplot}, we observe that $R_{\Phi}\mapsto \Fr(\Rsk,
R_{\Phi},R_M)$ is a non-increasing function. This is because as more
public information is made available to Bob, with all else fixed, the
key leakage rate increases, resulting in a smaller secrecy
exponent. The function $R_{M}\mapsto \Fr(\Rsk, R_{\Phi},R_M)$ is
non-decreasing because as Alice increases the use of her private
randomness through a larger $R_M$, she can conceal more of the key
from Eve. Finally, $\Rsk\mapsto \Fr(\Rsk, R_{\Phi},R_M)$ is
non-increasing because $\Rsk$ can be interpreted as the residual
source of secrecy that can be generated by Alice and Bob while keeping
Eve ignorant of the key generated.

\begin{figure}
  \centering
  \includegraphics[width=0.5\textwidth]{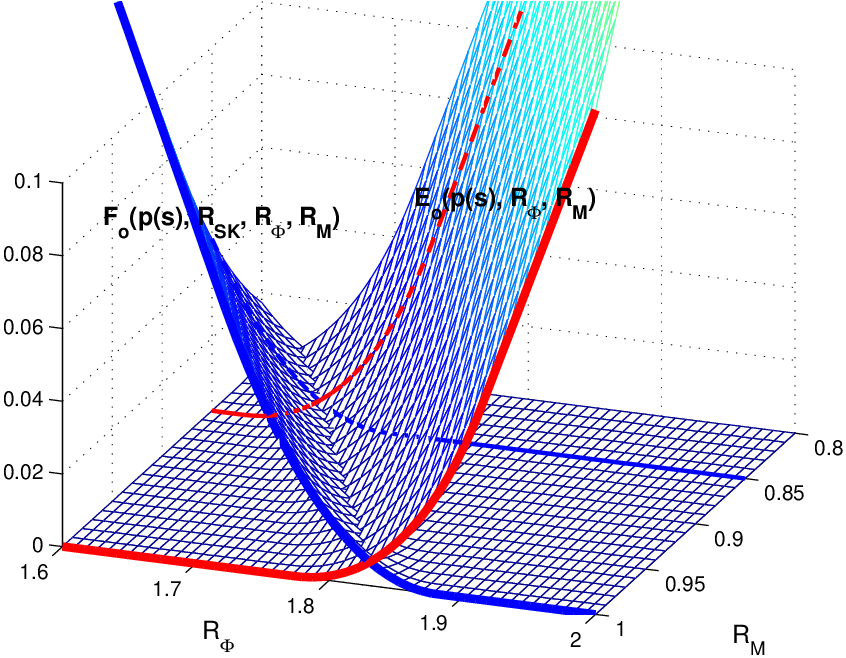}
  \caption{Plot of the reliability exponent $\Eo$ and secrecy exponent
    $\Fo$ for a fixed input distribution $p(s)= \randBern{0.5}$ with
    $\Rsk=0.01$. The exponents for two different values of $R_M$ are
    shown.}
  \label{fig:3d}
\end{figure}

\begin{figure}
  \centering
  \includegraphics[width=0.5\textwidth]{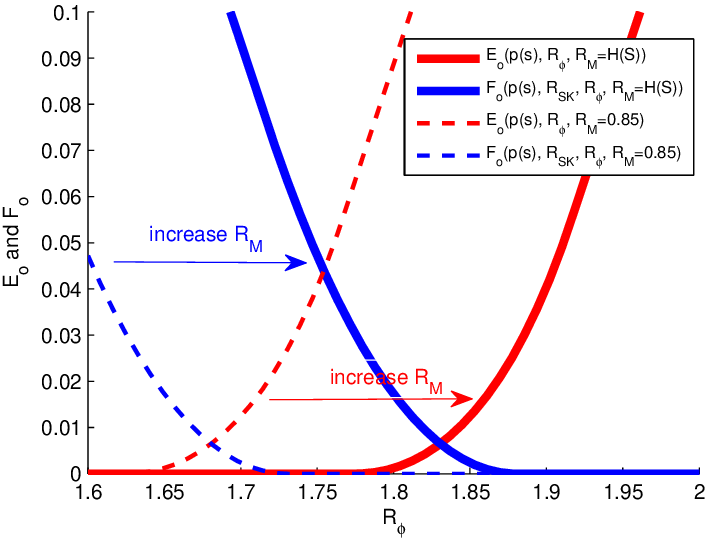}
  \caption{Two-dimensional visualization of Fig.~\ref{fig:3d}. The
    thick solid lines correspond to $R_M = H(S)=1$ and the thin dashed
    lines correspond to $R_M = 0.85$. }
  \label{fig:slice2d}
\end{figure}
 In Fig.~\ref{fig:3d}, we plot the exponents as a function of $R_\Phi$
 and $R_M$ for $\Rsk =0.01$. The input distribution $p(s)$ is kept
 fixed. Note that there is a non-empty region in the $(R_\Phi,R_M)$
 plane for which {\em both} exponents are positive, indicating that
 $\Rsk=0.01$ is strongly achievable. For clarity, we also present a
 two-dimensional visualization in Fig.~\ref{fig:slice2d} which helps
 to show the utility of our sender-excited model.  We observe the
 following: Suppose we want to have a secret key rate of $\Rsk=0.01$
 and that the public message rate must be limited to, say, $R_\Phi\leq
 1.68$ due to system constraints. Then by simply adopting a source
 emulation strategy, $R_M = H(S)=1$ (i.e., case (III) of
 Section~\ref{sec:connect_to_previous}), and the reliability exponent
 is zero even though the secrecy exponent is high.  The reliability
 and secrecy exponents for this choice of parameters is plotted with
 the thick solid lines.  Thus, we {\em cannot achieve} the key rate of
 $\Rsk=0.01$ with the fixed input distribution $p(s)$. However, our
 model affords us the flexibility to tune $R_M$.  If, for instance, we
 reduce it to $R_M =0.85$ while keeping $R_{\Phi} = 1.68$ we tradeoff
 a reduction in the secrecy exponent for an increase in the
 reliability exponent.  With this new choice of $R_M$ {\em both}
 exponents will be positive and the key rate $\Rsk=0.01$ is (strongly)
 achieved with the same fixed $p(s)$.  The exponents for this choice
 of parameters are plotted by the thin dashed lines.

\section{Proofs of   Results in Section \ref{sec:capacity_result}}
\label{sec:proof_capacity}

\subsection{Proof of Converse of Proposition~\ref{thm:capacity}} \label{sec:conv1}

We start with a lemma  \cite[Lemma 4.1]{Ahlswede_Csiszar93}, which is a consequence of the Csisz\'{a}r sum identity~\cite[Ch.\ 2]{ElGamal_Kim_LNIT}.
\begin{lemma} \label{lemma:Csiszar_identity}
    The following equality  holds for arbitrary random variables $K, \Phi, Y^n, Z^n$:
    \begin{align*}
        & I(K;Y^n|\Phi) - I(K;Z^n|\Phi) \\
        &\quad = \sum_{i=1}^n I(K;Y_i|Y^{i-1},Z_{i+1}^n, \Phi) - I(K;Z_i|Y^{i-1},Z_{i+1}^n, \Phi) \ .
    \end{align*}
\end{lemma}

\begin{proof}[Proof of Converse of Proposition~\ref{thm:capacity}]
Fix any sequence of $(2^{nR_M}, 2^{nR_\Phi}, n,\Gamma)$ codes  per  Section~\ref{sec:protocol}. Let $R_{\mathrm{SK}}$ be any $\Gamma$-weakly achievable rate per Definition~\ref{def:wach}.
Consider,
\begin{align}
    nR_{\mathrm{SK}}  &\le  I(\ka; \xb^n, \pmsg) + n\epsilon_n \label{eqn:fano} \\
        &\le I(\ka; \xb^n, \pmsg) - I(\ka; \xe^n, \pmsg) + 2n\epsilon_n \label{eqn:secrecy}\\
        &= I(\ka; \xb^n | \pmsg) - I(\ka; \xe^n | \pmsg) + 2n\epsilon_n \nonumber\\
        &= \sum_{i=1}^n I(\ka; \xb_i | \xb^{i-1}, \xe_{i+1}^n, \pmsg) \nonumber\\
        &\hspace{6ex} - I(\ka; \xe_i | \xb^{i-1}, \xe_{i+1}^n, \pmsg) + 2n\epsilon_n \label{eqn:conv1}
\end{align}
where \eqref{eqn:fano} is due to Fano's inequality ($\epsilon_n \to 0$ as $n\to\infty$), \eqref{eqn:secrecy} is due to the secrecy condition in~\eqref{equ:secrecy_condition} and~\eqref{eqn:conv1} by applying Lemma~\ref{lemma:Csiszar_identity}. Now we make the following identifications of the  auxiliary random variables
\begin{align}
W_i\triangleq (\xb^{i-1}, \xe_{i+1}^n, \pmsg)  ,\quad \mbox{and}\quad U_i \triangleq  (\ka, W_i)  \ .  \label{eqn:aux}
\end{align}
 As can be readily verified, the chosen variables $W_i$ and $U_i$ satisfy the Markov condition
\begin{align*}
 W_i   - U_i - ( S_i , X_i) - ( \xb_i , \xe_i)
\end{align*}
as required by~\eqref{eqn:joint_dist}.   Note that since $\KA$ and $\Phi$  (random variables contained in our identifications in $W_i$ and $U_i$ in \eqref{eqn:aux}) are both functions of $(M, X^n)$ (see   Section \ref{sec:problem_setting}), $S_i$ by itself does not separate $(X_i, Y_i, Z_i)$ from $W_i$ and $U_i$.  However, the separation {\em does} hold when $(S_i, X_i)$ are grouped together by the discrete memoryless nature of the channel $p(x,y,z|s)$. Substituting the choice of auxiliary random variables in \eqref{eqn:aux} into \eqref{eqn:conv1} yields,
\begin{align*}
nR_{\mathrm{SK}}&\le   \sum_{i=1}^n I(\ka; \xb_i | W_i) - I(\ka; \xe_i | W_i) + 2n\epsilon_n \\
&=\sum_{i=1}^n I(\ka, W_i; \xb_i | W_i) - I(\ka, W_i; \xe_i | W_i) + 2n\epsilon_n \\
        &= \sum_{i=1}^n I(U_i ; \xb_i | W_i) - I(U_i ; \xe_i | W_i) + 2n\epsilon_n \ .
\end{align*}
Now, introduce the time-sharing random variable $Q$ with uniform distribution $\rvP(Q =i) = 1/n$ for all $i\in [1:n]$ and independent of $(W^n,U^n,S^n,X^n,Y^n,Z^n)$. Define the random variables $U \triangleq (U_Q,Q)$, $W \triangleq (W_Q,Q)$, $S\triangleq S_Q$, $X\triangleq X_Q$, $Y\triangleq Y_Q$ and $Z\triangleq Z_Q$. Then, we have
\begin{align}
R_{\mathrm{SK}} &  \le\sum_{q=1}^n\rvP(Q =q)\big[  I(U_q ; Y_q|W_q) - I(U_q ; Z_q|W_q)\big] + 2 \epsilon_n \nnum\\
&= I(U_Q ; Y_Q|W_Q,Q) - I(U_Q ; Z_Q|W_Q,Q) + 2 \epsilon_n\nnum\\
&= I(U_Q, Q ; Y_Q|W_Q,Q) - I(U_Q ,Q ; Z_Q|W_Q,Q)+ 2 \epsilon_n\nnum \\
&= I(U  ; Y |W ) - I(U  ; Z |W )+2 \epsilon_n \label{eqn:RSK_bound}  \ .
\end{align}
Note also that since $S^n$ satisfies the  almost sure cost constraint in \eqref{equ:cost_constraint}, $\frac{1}{n}\sum_{i=1}^n \rvE [\Lambda(S_i) ]\le\Gamma$ holds. This implies from the definition of $Q$ and  $S$ that $\rvE [\Lambda(S)] = \rvE_Q \{\rvE[\Lambda(S_Q)\, | \, Q] \}  \le\Gamma$. Thus to remove the dependence on the code, we maximize~\eqref{eqn:RSK_bound} over all joint distributions that satisfy~\eqref{eqn:joint_dist} and $\rvE [\Lambda(S)]\le\Gamma$, i.e.,
\begin{align*}
R_{\mathrm{SK}}\le\max_{\substack{ W-U-(X,S)-(Y,Z) \\ \rvE[\Lambda(S)]\le\Gamma}}I(U  ; Y |W ) - I(U  ; Z |W )+2 \epsilon_n  \ .
\end{align*}
Taking $n\to\infty$ completes the proof of the converse.\end{proof}

\subsection{Proof of Proposition~\ref{thm:upper_bound}} \label{sec:loose_ub}
\begin{proof}
We   prove the upper bound in \eqref{equ:upper_bound}. Consider the inequalities:
    \begin{align}
        n \Rsk &\le  I(\ka; \xb^n, \pmsg) + n\epsilon_n    \label{eqn:fano2}\\
        &\le I(\ka; \xb^n, \pmsg, \xe^n) + n\epsilon_n \nonumber\\
         &= I(\ka; \xb^n| \pmsg, \xe^n)+I ( \ka;\pmsg, \xe^n) + n\epsilon_n \nonumber\\
               &\le I(\ka; \xb^n| \pmsg, \xe^n) + 2 n\epsilon_n\label{eqn:secre} \\
               &\leq I(\ka, \pmsg; \xb^n| \xe^n) + 2 n\epsilon_n   \ , \label{eqn:upper_bd_1}
               \end{align}
   where \eqref{eqn:fano2} follows Fano's inequality and \eqref{eqn:secre} is due to the secrecy condition \eqref{equ:secrecy_condition}. Continuing from \eqref{eqn:upper_bd_1}, we have
           \begin{align}
             n \Rsk    &\le  I(X^n, M ; \xb^n| \xe^n)   + 2 n\epsilon_n \label{eqn:func}\\
               &= I(X^n; \xb^n| \xe^n)+ I(M;Y^n|X^n,Z^n) + 2 n\epsilon_n \nonumber\\
               &\le I(X^n; \xb^n| \xe^n)+ I(S^n;Y^n|X^n,Z^n) + 2 n\epsilon_n \label{eqn:depends}\\
               &= I(\xin^n; \xb^n| \xe^n) \! +\! I(\xa^n; \xb^n| \xin^n, \xe^n) + 2 n\epsilon_n    ,  \label{eqn:twoterms}
    \end{align}
    where  \eqref{eqn:func} follows  because $(K_{\mathrm{A}},\pmsg)$ is a function of  $(X^n, M )$ and \eqref{eqn:depends} follows  because the channel only depends on $S^n$ so $M-S^n -(X^n, Y^n,Z^n)$.\footnote{In fact, \eqref{eqn:depends}  holds with equality because $S^n=S^n(M)$ in addition to the stated Markov relationship.}  Now the first term \eqref{eqn:twoterms} can be   upper bounded as  follows
        \begin{align}
            &I(\xin^n; \xb^n| \xe^n) = H(\xb^n| \xe^n) - H(\xb^n| \xin^n, \xe^n) \nonumber\\
            &= \sum_{i=1}^n H(\xb_i|\xb^{i-1}, \xe^n) - H(\xb_i| \xb^{i-1}, \xin^n, \xe^n) \nonumber\\
            &\leq \sum_{i=1}^n H(\xb_i| \xe_i) - H(\xb_i| \xin_i, \xe_i)  = \sum_{i=1}^n I(\xin_i; \xb_i| \xe_i)\ ,  \label{eqn:firstterm}
        \end{align}
        where the inequality follows by conditioning reduces entropy and the Markov chain $(Y^{i-1}, Z^{n\setminus i},S^{n\setminus i})-(S_i,Z_i)-Y_i$. The second term in \eqref{eqn:twoterms}  can be written as a sum:
\begin{equation}
I(\xa^n; \xb^n| \xin^n, \xe^n) = \sum_{i=1}^n I(\xa_i; \xb_i| \xin_i, \xe_i)    \label{eqn:secterm}
\end{equation}
         because the channel $p(x,y,z|s)$ is memoryless. Substituting~\eqref{eqn:firstterm} and \eqref{eqn:secterm} into \eqref{eqn:twoterms} yields
\begin{align}
     n \Rsk & \le  \sum_{i=1}^n I(\xin_i; \xb_i| \xe_i) + I(\xa_i; \xb_i| \xin_i, \xe_i) + 2 n\epsilon_n \nnum \\
              &= \sum_{i=1}^n I(\xa_i,\xin_i; \xb_i| \xe_i) + 2 n\epsilon_n\  . \label{eqn:sumterm} 
\end{align}
The proof can be completed using the time-sharing technique in  the converse proof of Proposition~\ref{thm:capacity}. \end{proof} 

\section{Proofs of Results in Section \ref{sec:exponent_result}}
\label{sec:proof_exponent}
In this section, we provide the proof of Theorem~\ref{thm:inner_bound_triple} on the  capacity-reliability-secrecy region. This section will be split into three subsections: In the first subsection, we collect some  relevant definitions and describe the coding scheme. The second  and third subsections contain the proofs of the achievability (lower bounds) of the reliability  and secrecy exponents  respectively.  This proves the achievability of the region $ \tilde{\calR}(p(s), \Rpub, R_M)$  defined in~\eqref{eqn:R_ps}.

\subsection{Definitions and Coding Scheme} \label{sec:defs}
We start  with some definitions to describe the generation of the codewords $s^n(m)$, the key  and the public message generation procedures.
\begin{definition}[Random code] \label{def:random_code}
A $(2^{nR_M},n)$ {\em random code generated according to} $p(s)$ is a random subset of $\calS^n$ which contains length-$n$ sequences $s^n(m),m\in [1:2^{nR_M}]$ where each sequence $s^n(m)$, called a {\em codeword},  is drawn according to the pmf $\prod_{i=1}^n p(s_i)$.
\end{definition}
Note that we do not place any cost  constraints on $p(s)$ because we assume that $\Gamma=\infty$ in Section~\ref{sec:exponent_result}.

\begin{definition}[Random binning  function \cite{gallager76}] \label{def:rbf}
A $2^{nR}$ {\em random binning function  for an alphabet} $\calU$ is a random map\footnote{More precisely, $\psi(b|u)$ is a matrix of conditional probabilities.    }  $\psi: u\in \calU \to    b \in [1:2^{nR}]$ that satisfies the following properties:
\begin{itemize}
\item {\bf Uniformity}:   Each element $u\in\calU$ is independently and uniformly assigned to an element of $[1:2^{nR}]$.
\item  {\bf Pairwise Independence}:  Each pair of different $u,u'\in\calU$ is mapped $u\mapsto b$, $u'\mapsto b'$ with probability $2^{-2nR}$ for each pair of elements $b,b'\in [1:2^{nR}]$ (not necessarily different).
\item The random map $\psi$ is independent of the random code generation process as per Definition~\ref{def:random_code}. More precisely,
\begin{align*}
\rvP(\{S^n=s^n\} \cap \{\psi(u) \!= b \}) \! = \rvP( S^n\!=\!s^n ) \rvP( \psi(u) = b   )
\end{align*}
\end{itemize}
\end{definition}



We now introduce the notion of a random binning code  for the secret key generation protocol (See Section~\ref{sec:protocol}).
\begin{definition}[Random binning secret key  code] \label{def:random_bin}
A  $(2^{n\Rsk}, 2^{nR_M}, 2^{nR_{\Phi}},n)$ {\em random binning secret key code} is a $( 2^{nR_M}, 2^{nR_{\Phi}},n)$ code for the secret key generation protocol  in which the public message and key are generated via two   independent random binning functions:
\begin{align}
 \phi &: \calM\times\calX^n  \to \alppmsg=[1:2^{nR_{\Phi}}] \label{eqn:phi_def}  \\
 \kA &:  \calM\times\calX^n  \to \calK=[1:2^{n\Rsk}]   \label{eqn:ka_def}  \ .
\end{align}
\end{definition}
More precisely, note from \eqref{eqn:phi_def} that  $\phi$ is a $2^{nR_{\Phi}}$ random binning function for alphabet $\calM\times\calX^n $ and from~\eqref{eqn:ka_def} that $\kA$ is a $2^{n\Rsk}$ random binning function for alphabet $\calM\times\calX^n $.   \\

\noindent {\bf Codebook Generation and Encoding}:  Fix $p(s)$.  We use a  $(2^{n\Rsk}, 2^{nR_M}, 2^{nR_{\Phi}},n)$  random binning secret key code in which the codewords $s^n(m),m\in\calM$ belong to a $(2^{nR_M}, n)$ random code  generated according to  $p(s)$.   The codewords and bin assignments are revealed to all parties before communication starts.   We emphasize that by construction, this $(2^{n\Rsk}, 2^{nR_M}, 2^{nR_{\Phi}},n)$  code is a $(2^{nR_M}, 2^{nR_{\Phi}},  n)$  code (in the sense of Section~\ref{sec:protocol} with $\Gamma=\infty$) such that secret key rate $\Rsk$ is    achievable. This is because  $\KA$ is uniformly distributed on  $[1:2^{n\Rsk}]$ so~\eqref{eqn:sk_rate} is satisfied.

By the definition of $ \tilde{\calR}(p(s), \Rpub, R_M)$ in~\eqref{eqn:R_ps}, it suffices to show the following two assertions hold true for any $p(s)$:
\begin{align}
&\liminf_{n\to\infty}  -\frac{1}{n} \log \rvP(\KA\ne\KB)\ge\Eo(p(s), \Rpub, R_M)  , \nnum\\
&\liminf_{n\to\infty} -\frac{1}{n} \log I( \KA; \xe^n,\Phi) \ge\Fo(p(s), \Rsk, \Rpub, R_M).\nnum
\end{align}
This is what we prove in the next two subsections.

\subsection{Proof for the Reliability Exponent  } \label{sec:reliable_prf}
In this section, we will prove that $E_{\mathrm{o}}$ is an achievable reliability exponent.  Recall that Bob has access to his channel output $y^n\in\calY^n$ and the public message $\phi\in\alppmsg$, which was generated by Alice in accordance to the random binning function in~\eqref{eqn:ka_def}.  In order to analyze the error event that Bob's key does not match Alice's
\begin{equation}
\calE_{\mathrm{key}}\triangleq \{\KA\ne\KB\} \ , \label{eqn:eventK}
\end{equation}
we   stipulate that Bob decodes {\em both} Alice's received sequence $x^n\in\calX^n$ and Alice's source of randomness $m\in\calM$.

We restate the ML-MAP decoding rule in~\eqref{eqn:mlmap1}: Given $(y^n, \phi)$,  Bob declares that $m$ is the message selected by Alice and  $x^n$ is the sequence sent to Alice if the public message bin index of $(m,x^n)$ agrees with $\phi$, i.e.,
\begin{equation}
\phi( m,x^n) = \phi  \label{eqn:phi_bin}
\end{equation}
and the probabilities satisfy
\begin{align}
p(y^n|s^n(m) &  ) p (x^n| y^n,s^n(m) )\ge  \nnum \\
& \qquad \qquad  p(y^n|s^n(\tilm))p(\tilx^n|  y^n,s^n(\tilm)) \label{eqn:mlmap}
\end{align}
for all other pairs $(\tilm,\tilx^n)$ such that $\phi(\tilm,\tilx^n) = \phi$.  As mentioned previously, this is a hybrid of an ML  and  an MAP  rule.   Observe that  if we were just to maximize $p(y^n|s^n(m))$ over $m$, this would correspond to a pure ML decoding rule for the channel $p(y|s)$ as in \cite[Sec.\ 5.6]{gallagerIT}. If instead we maximize $p (x^n| y^n,s^n(m) )$ over $x^n$ given $m$ is known, this would correspond to a pure MAP decoder for the source $x^n$ given side information $(m,y^n)$ as in \cite{gallager76}.

By analyzing the ML-MAP decoder, we now upper bound the probability of event $\calE_{\mathrm{key}}$ of the ensemble random binning secret key code $\scC$, \ie, $\rvP(\calE_{\mathrm{key}}) \triangleq \rvE_{\scC}[\rvP(\calE_{\mathrm{key}}  |\scC) ] =  \sum_{\calC}p(\calC) \rvP(\calE_{\mathrm{key}}  |\scC\!=\!\calC)$. Throughout, we use the notation $\scC$ to denote the random code (a random variable) and $\calC$ to denote a specific code.   Define  the error event  that  Bob decodes either $M$ or $X^n$ incorrectly
\begin{equation}
\calE\triangleq\{ (\hatM,\hatX^n) \ne (M,X^n)\} \ . \label{eqn:eventE}
\end{equation}
Clearly, $\calE_{\mathrm{key}} \subset\calE$.  Thus, an upper bound for $\rvP(\calE)$ also serves as an upper bound for $\rvP(\calE_{\mathrm{key}})$. Similarly, a lower bound for the exponent of $\rvP(\calE)$ is also a lower  bound for the exponent of $\rvP(\calE_{\mathrm{key}})$. In the interest of tractability, we upper bound $\rvP(\calE)$ [instead of $\rvP(\calE_{\mathrm{key}})$]   when the ML-MAP decoder described  in~\eqref{eqn:phi_bin} and~\eqref{eqn:mlmap} is used. In order to bound $\rvP(\calE)$, we decompose $\calE$ into the following three disjoint error events:
\begin{align}
\calE_1 \triangleq \{ \hatM = M, \hatX^n \ne X^n\}  \label{eqn:E1} \\
\calE_2 \triangleq \{ \hatM \ne  M, \hatX^n = X^n\} \label{eqn:E2} \\
\calE_3 \triangleq \{ \hatM  \ne M, \hatX^n \ne X^n\}  \label{eqn:E3} 
\end{align}
Note that the error exponent is the minimum of the exponents for $\rvP(\calE_1)$, $\rvP(\calE_2)$  and $\rvP(\calE_3)$. In the following, we only provide a detailed derivation for $\rvP(\calE_3)$ as it is the most interesting and unconventional. We note that for $\calE_1$,  if $M=m$, $p(\hatx^n|y^n, s^n(m))\ge   p(x^n|y^n, s^n(m))$ (the MAP decoding part) so this analysis   parallels that by Gallager for Slepian-Wolf coding~\cite{gallager76} (reconstructing $X^n$ given side information $(Y^n, S^n(M))$ and $M$ is decoded correctly). Thus, we have
\begin{align}
&\liminf_{n\to\infty}  -\frac{1}{n}\log \rvP(\calE_1) \ge \nnum\\
 &  \rho R_\Phi - \log \sum_{s,y} p(s) p(y|s) \left(\sum_x p(x|y,s)^{1/(1+\rho)}\right)^{1+\rho}. \label{eqn:PE1}
\end{align}
Similarly for $\calE_2$, we have that  $p(x^n, y^n|s^n(\hatm))\ge p(x^n, y^n|s^n( m))$ (Bayes rule)  so this is simply the error in ML decoding for channel coding with vector output $(X,Y)$ and input $S$. Consequently, from Gallager's book~\cite[Sec.~5.6]{gallagerIT},
\begin{align}
&\liminf_{n\to\infty}  -\frac{1}{n}\log \rvP(\calE_2) \ge \nnum\\
 &   \rho (R_\Phi-R_M) - \log \sum_{s}  \left(\sum_{x,y}  p(s) p(x,y|s)^{1/(1+\rho)}\right)^{1+\rho}.\label{eqn:PE2}
\end{align}
Here we note that there are $\doteq 2^{nR_M}$ sounding sequences $s^n(m)$ but by \eqref{eqn:phi_bin}, we  search within a particular bin indexed by $\phi$ so effectively, there are only $\doteq 2^{n(R_M-R_\Phi)}$   sounding sequences explaining the leading term in \eqref{eqn:PE2}.

Now, we analyze $\rvP(\calE_3)$ in detail. Consider the probability of error given that  $m$ is the message  sent, $s^n(m)$ represents  the ensemble of codewords associated to $m$ (by the random codebook construction in Definition~\ref{def:random_code}), $x^n$ is Alice's received sequence and $y^n$ is Bob's received sequence. That is,  consider
\begin{align}
\rvP(&  \calE_3|y^n  , s^n(m), m, x^n)   \nnum\\
&= \rvP\left( \bigcup_{\hatm\ne m ,s^n(\hatm) , \hatx^n\ne x^n }   \calA(  s^n(\hatm), \hatm,  \hatx^n  ) \right)   .\label{eqn:error1}
 \end{align}
In the above error probability,   $\calA(  s^n(\hatm), \hatm,  \hatx^n  )$  is defined as the  error   event  that the message $\hatm\ne m$, codeword $s^n(\hatm)$ and Alice's sequence $\hatx^n\ne x^n$ are selected in such a way that their ML-MAP objective value is higher than that of the true parameters $(m,s^n(m), x^n)$, i.e.,  that $ p(y^n|s^n(\hatm)) p(\hatx^n|y^n, s^n(\hatm))\ge p(y^n|s^n(m)) p(x^n|y^n, s^n(m))$ and also that $\phi(\hatm, \hatx^n)=\phi(m,x^n)$.  Note in \eqref{eqn:error1} that the error event is averaged over all incorrect codewords $s^n(\hatm)$ due to the random codebook construction (Definition~\ref{def:random_code}).    Now recall the assumption that  the binning process is pairwise independent and also independent of the inputs (Definition~\ref{def:rbf}). More precisely,
\begin{align}
\rvP( & \{S^n=s^n(\hatm)\}\cap \{\phi(m,x^n) = \phi(\hatm, \hatx^n) \})   \nnum\\
&= \rvP(  S^n=s^n(\hatm) ) \rvP(  \phi(m,x^n) = \phi(\hatm, \hatx^n)  )  \nnum\\
&= p(s^n(\hatm)) \sum_{\phi\in\alppmsg} \frac{1}{|\alppmsg|^2} = \frac{p(s^n(\hatm))}{|\alppmsg|} \ . \label{eqn:indep}
\end{align}
Let $\mathbf{1}_{\calB}$ be the indicator variable of the set $\calB$. By using the definition of  $\calA( s^n(\hatm),\hatm,\hatx^n)$ and \eqref{eqn:indep}, we can upper bound the probability of $\calA( s^n(\hatm),\hatm,\hatx^n)$ as follows:
\begin{align}
& \rvP(\calA( s^n(\hatm),\hatm,\hatx^n)) \nnum\\
&=  \frac{p(s^n(\hatm))}{|\alppmsg|} \mathbf{1}_{\{  p(\hatx^n, y^n| s^n(\hatm))\ge  p(x^n,y^n| s^n(m)) \}}  \nnum\\
&\le \frac{p(s^n(\hatm))}{|\alppmsg|} \left( \frac{ p(y^n|s^n(\hatm)) p(\hatx^n|y^n, s^n(\hatm))}{ p(y^n|s^n(m)) p(x^n|y^n, s^n(m))} \right)^t \ ,  \nnum
\end{align}
for all $t>0$, where the inequality follows because $\mathbf{1}_{\{a\ge b\}}\le (\frac{a}{b})^t$ for all $t>0$. Let $\rho\in [0,1]$. By applying the inequality $\rvP\left(\cup_{t=1}^T \calA_t\right)\le [ \sum_{t=1}^T \rvP \left( \calA_t\right) ]^{\rho}$  \cite[pp.\ 136]{gallagerIT}  to~\eqref{eqn:error1}, we have
\begin{align}
&\rvP(\calE_3|y^n , s^n(m), m, x^n)   \nnum\\
&\le \Bigg[  \sum_{\hatm\ne m ,s^n(\hatm) , \hatx^n\ne x^n } \frac{p(s^n(\hatm) )}{|\alppmsg|}  \times \ldots \nnum\\
 & \qquad \qquad \times\left( \frac{ p(y^n|s^n(\hatm)) p(\hatx^n|y^n, s^n(\hatm))}{ p(y^n|s^n(m)) p(x^n|y^n, s^n(m))} \right)^t \Bigg]^{\rho} \label{eqn:PEy}
 \end{align}
for any $\rho \in [0,1]$ and $t>0$.  Now consider the error probability $\rvP(\calE_3|M=m)$ given message $m$ is chosen by Alice, i.e., $\{M=m\}$ occurs. To bound this error probability, we average over all codewords $s^n(m)$, all observed sequences $y^n$ and all possible sequences received by Alice $x^n$, i.e.,
\begin{align}
& \rvP(\calE_3  |   m )  = \sum_{y^n}\sum_{s^n(m)} p(y^n|s^n(m)) p(s^n(m)) \times\ldots\nnum\\
&\qquad\times  \sum_{x^n} p(x^n|y^n, s^n(m))\rvP(\calE_3|y^n , s^n(m), m, x^n) \ .   \label{eqn:PcalEm}
\end{align}
We now substitute the upper bound in~\eqref{eqn:PEy} into~\eqref{eqn:PcalEm}. Pulling out   $ p(x^n|y^n, s^n(m))$ from the innermost term in \eqref{eqn:PEy} (since it does not depend on $\hatm$, $s^n(\hatm)$ and $\hatx^n$), we see that $\rvP(\calE_3|   m )$ can be upper bounded as
\begin{align}
 \rvP(&\calE_3|   m ) \le  |\alppmsg|^{-\rho} \sum_{y^n}\sum_{s^n(m)}  p(y^n|s^n(m))  p(s^n(m)) \times\ldots\nnum\\
&    \times   \sum_{x^n} p(x^n|y^n,s^n(m))^{1-\rho t} \Bigg[ \sum_{\hatm\ne m}\sum_{s^n(\hatm)} p(s^n(\hatm)) \times\ldots\nnum\\
&   \times \left(\frac{p(y^n|s^n(\hatm))}{p(y^n|s^n(m))}\right)^t \sum_{\hatx^n\ne x^n} p(\hatx^n|y^n, s^n(\hatm))^t \Bigg]^{\rho}  \nnum   \\
&= |\alppmsg|^{-\rho}  (|\calM|-1)^{\rho} \sum_{y^n} \Psi_1(y^n, \rho, t) \Psi_2(y^n, \rho, t) \ ,  \label{eqn:before_dummy}
\end{align}
where the functions  $\Psi_1(y^n, \rho, t)$ and  $\Psi_2(y^n, \rho, t)$ are defined as  follows:
\begin{align}
\Psi_1(y^n, \rho, t) &\triangleq    \sum_{s^n(m)} p(s^n(m)) p(y^n|s^n(m))^{1-\rho t}  \times\ldots\nnum\\
&\qquad \qquad     \times \sum_{x^n} p(x^n|y^n, s^n(m))^{1-\rho t} \nnum \\
 \Psi_2(y^n, \rho, t)& \triangleq \Bigg[ \sum_{s^n(\hatm)} p(s^n(\hatm)) p(y^n|s^n(\hatm))^{t} \times\ldots\nnum\\
 &\qquad \qquad   \times  \sum_{\hatx^n} p(\hatx^n|y^n, s^n(\hatm))^{ t}\Bigg]^{\rho} \ .  \nnum
\end{align}
Equation~\eqref{eqn:before_dummy} follows because $\hatm$  in the line above is a dummy variable that can take on exactly $|\calM|-1$ values and for each $\hatm$, we generate codewords $s^n(\hatm)$ in the {\em same way} in the random coding construction.   Now notice that if we set $t= 1/( 1+\rho)$, then $$\Psi_2(y^n, \rho,1/( 1+\rho) ) = \Psi_1 (y^n, \rho,1/( 1+\rho) )^{\rho}$$ because $\hatx^n$ and $\hatm$ in the definition of $\Psi_2$ are dummy variables. As such,   $\rvP(\calE_3  |   m )$ can be bounded as
\begin{equation}
\rvP(\calE_3  |   m ) \le |\alppmsg|^{-\rho} |\calM|^{\rho} \sum_{y^n} \Psi_3(y^n, \rho) \ ,  \label{eqn:dummy}
\end{equation}
where the function $\Psi_3(y^n, \rho)$ is defined as
\begin{align}
\Psi_3(y^n, \rho) &\triangleq \Bigg[ \sum_{s^n(m)} p(s^n(m)) p(y^n|s^n(m))^{1/(1+\rho)} \times\ldots \nnum\\
&\qquad  \sum_{x^n} p(x^n|y^n, s^n(m))^{1/(1+\rho)}\Bigg]^{1+\rho}  \ .  \nnum
\end{align}
Now, we recall the DMS and DMBC assumptions, i.e., that
\begin{align}
p(s^n(m))&=\prod_{i=1}^n p(s_i(m)) \ ,   \nnum \\ p(x^n, y^n|s^n(m)) &= \prod_{ i=1}^n p(x_i,y_i|s_i(m)) \ .  \nnum
\end{align}
As a result, $\Psi_3(y^n, \rho)$   simplifies to
\begin{align}
\Psi_3(y^n, \rho)& = \Bigg[ \prod_{i=1}^n \sum_{s_i(m)} p(s_i(m))p(y_i|s_i(m))^{1/(1+\rho)} \times\ldots \nnum\\
&\qquad \times  \sum_{x_i}  p(x_i|y_i,s_i(m))^{1/(1+\rho)} \Bigg]^{ 1+\rho} \ ,  \nnum
\end{align}
and the sum in \eqref{eqn:dummy} can    be written as a product of single-letterized  terms:
\begin{equation}
 \sum_{y^n} \Psi_3(y^n, \rho)  =  \prod_{i=1}^n \sum_{y_i}  \Psi_4(y_i,\rho)  \ ,  \label{eqn:Psi4}
\end{equation}
where the function $ \Psi_4(y,\rho)$ is defined as
\begin{align}
\Psi_4( y, \rho) &\triangleq   \left[\sum_s p(s) p(y|s)^{1/(1+\rho)} \sum_x p(x|y,s )^{1/(1+\rho)}\right]^{1+\rho}   . \nnum
\end{align}
Because each of the codewords is generated identically, each of the terms in the product in~\eqref{eqn:Psi4} is also identical. Hence,
\begin{equation}
 \sum_{y^n} \Psi_3(y^n, \rho)   = \left[  \sum_{y}  \Psi_4(y,\rho)  \right]^n \  . \nnum
\end{equation}
Recall that $|\alppmsg|\doteq 2^{n R_{\Phi}}$ and $|\calM|\doteq 2^{nR_M}$. In addition, note that $\rvP(\calE_3)= \sum_{m'} p(m') \rvP(\calE_3|   m') = \rvP(\calE_3|   m)$ for every $m\in\calM$.  As such, taking the normalized logarithm and limit  inferior of~\eqref{eqn:dummy} yields
\begin{align}
 \liminf_{n\to\infty} -\frac{1}{n}\log\rvP(\calE_3)\ge  \rho(R_{\Phi}-R_M)  -\log   \sum_y \Psi_4( y, \rho)  \ .  \label{eqn:PE3}
\end{align}
Essentially, what we have done is to develop a ``hybrid'' of Gallager-style error exponents for channel and lossless source coding with side information.  Thus, an achievable error exponent when input distribution $p(s)$ is used is $E_{\mathrm{o}}^{(3)}(p(s), R_{\Phi},R_M)$  defined in \eqref{eqn:relaible_expo}.
The reliability exponent part of the theorem  is proved for the  random binning secret key code by combining the bounds for the exponents for $\rvP(\calE_1),\rvP(\calE_2)$ and $\rvP(\calE_3)$ in \eqref{eqn:PE1}, \eqref{eqn:PE2} and \eqref{eqn:PE3} respectively. \qed

\subsection{Proof for the Secrecy Exponent }
We now prove that the secrecy exponent is at least $F_{\mathrm{o}}$ using the same coding scheme.    We can  use steps  analogous to the proof of the direct part of Theorem~2 in~\cite{chou_it10} to  obtain  the following  bound on the key leakage  $I(\KA;Z^n, \Phi)$.
\begin{lemma} \label{lem:info_leak}
Define $c(\alpha)\triangleq \alpha^{-1} \log e$ for $0<\alpha\le 1$. The key leakage can be bounded as follows:
\begin{align}
&  I(\KA   ;Z^n, \Phi)=\rvE_{\scC}[I( \KA; Z^n,\Phi| \scC)]  \nnum\\
&  \le      c(\alpha) \, |\calK|^{\alpha} |\alppmsg|^{\alpha}  \sum_{z^n} p(z^n )  \sum_{m,x^n} p(m,x^n|z^n )^{1+\alpha}   , \label{eqn:upper_bd_mi2}
\end{align}
for all $0<\alpha\le 1$.
\end{lemma}
The  proof   is provided at the end for completeness.  Now we consider the inner sum  in~\eqref{eqn:upper_bd_mi2}. By introducing the input $s^n$ and by repeated applications of Bayes rule,
\begin{align}
&\sum_{m,x^n} p(m,x^n|z^n )^{1+\alpha}  \nnum\\
&=   \sum_{x^n} \sum_m \left[ \sum_{s^n} p(m,x^n, s^n|z^n) \right]^{1+\alpha} \nnum \\
&=   \sum_{x^n} \sum_m \left[ \sum_{s^n} \frac{p(m,x^n ,s^n,z^n)}{p(z^n)} \right]^{1+\alpha} \nnum \\
&=  \frac{1}{p(z^n)^{ 1+\alpha }}  \sum_{x^n} \sum_m \Theta_1 (m,x^n,z^n)^{1+\alpha} \label{eqn:markv1}   \\
&= \frac{1}{p(z^n)^{ 1+\alpha } |\calM|^{1+\alpha} }\sum_{x^n} \sum_m \Theta_2 (m,x^n,z^n)^{1+\alpha} \label{eqn:uniformity}
\end{align}
where the functions $\Theta_1 (m,x^n,z^n)$ and $\Theta_2 (m,x^n,z^n)$ are defined as
\begin{align}
\Theta_1 (m,x^n,z^n)&\triangleq  \sum_{s^n} p(m)p(s^n|m)p(z^n|s^n) p(x^n|s^n,z^n) \nnum  \\
\Theta_2 (m,x^n,z^n)&\triangleq   \sum_{s^n} p(s^n|m)p(z^n|s^n) p(x^n|s^n,z^n)  \ .  \nnum
\end{align}
Equation \eqref{eqn:markv1}   follows because $M-S^n-(X^n,Z^n)$ form a Markov chain so $p(z^n|s^n,m) =p(z^n|s^n)$ and $p(x^n|s^n,z^n,m) = p(x^n|s^n,z^n)$. Equation~\eqref{eqn:uniformity} follows from the uniformity of the messages $m$ in the message set $\calM$, i.e., that $p(m) = \frac{1}{|\calM|}$ for all $m\in\calM$.  We now  upper bound $\Theta_2 (m,x^n,z^n)^{1+\alpha}$. This is done using the following lemma.
\begin{lemma} \label{lem:convex_ineq}
Let $\{(\lambda_j,a_j)\}$ be a finite collection of non-negative numbers such that  $\sum_j \lambda_j =1$. Also, let $r\ge 1$. Then, the  following inequality holds
\begin{equation}
\Bigg(\sum_j \lambda_j a_j \Bigg)^r \le  \sum_j \lambda_j a_j^r \ . \nnum
\end{equation}
\end{lemma}
This can be proven by noticing that  $t\mapsto t^r$ is convex. We omit the details.  We now make the following identifications:  $a_{s^n} \equiv p(z^n|s^n )  p(x^n| s^n,  z^n)$, $\lambda_{s^n} \equiv  p(s^n| m)$ and $r\equiv 1+\alpha$ and apply Lemma~\ref{lem:convex_ineq} to  $\Theta_2 (m,x^n,z^n)^{1+\alpha}$. This yields the inequality
\begin{align}
\Theta_2 (m,x^n,z^n)^{1+\alpha}\le  \sum_{s^n} p(s^n|m) [p(z^n|s^n) p(x^n|s^n,z^n)]^{1+\alpha}. \label{eqn:theta2ineq}
\end{align}
On account of  \eqref{eqn:upper_bd_mi2}, \eqref{eqn:uniformity} and \eqref{eqn:theta2ineq}, we have
\begin{align}
  & \rvE_{\scC}[I(\KA; Z^n,\Phi| \scC)] \le  c(\alpha)\,  |\calK|^{\alpha} |\alppmsg|^{\alpha} |\calM|^{-(1+\alpha)} \times \ldots \nnum\\
 & \sum_{z^n} p(z^n)^{-\alpha} \sum_{ s^n,x^n,m} p(s^n| m)  \left[ p(z^n|s^n )  p(x^n| s^n,  z^n) \right]^{1+\alpha}    \nnum\\
&  = c(\alpha)\,  |\calK|^{\alpha} |\alppmsg|^{\alpha} |\calM|^{-(1+\alpha)} \times \ldots \nnum\\
& \sum_{s^n,x^n,z^n}  \sum_m p( s^n,x^n,z^n  |m)  \left[\frac{p(z^n|s^n)}{p(z^n)} p(x^n|s^n,z^n) \right]^{\alpha} ,   \nnum
\end{align}
where  the final equality   follows because $p( s^n,x^n,z^n  |m)  = p(s^n|m)p(z^n|s^n )  p(x^n| s^n,  z^n)$ by the Markov chain $M-S^n-(X^n,Z^n)$. Now, pulling  the  $p(m)=\frac{1}{|\calM|}$ term into the sum, we get
\begin{align}
 & \rvE_{\scC}[I(\KA; Z^n,\Phi| \scC)] \le c(\alpha)\,  |\calK|^{\alpha} |\alppmsg|^{\alpha}  |\calM|^{-\alpha}  \times \ldots \nnum\\
 &\sum_{ s^n,x^n,z^n  } \sum_m p(  s^n,x^n,z^n  |m) p(m)  \left[\frac{p(z^n|s^n)}{p(z^n)} p(x^n|s^n,z^n) \right]^{\alpha}   \nnum\\
&= c(\alpha) \,  |\calK|^{\alpha} |\alppmsg|^{\alpha}  |\calM|^{-\alpha}  \sum_{ s^n,x^n,z^n } \Upsilon(s^n,x^n,z^n,\alpha)  \ , \nnum
\end{align}
where the function $\Upsilon(s^n,x^n,z^n,\alpha) $ is defined as
$$
\Upsilon( s^n,x^n,z^n,\alpha) \triangleq    p(s^n,x^n,z^n  )  \left[\frac{p(z^n|s^n)}{p(z^n)} p(x^n|s^n,z^n) \right]^{\alpha}   \ .
$$
Now, recall that (i) the input  $S^n$ is a DMS when averaged over all codebooks and all messages $m\in\calM$ (because the generation of the codewords $s^n(m),m\in\calM$ is done {\em identically}) and (ii) $p(x,y,z|s)$ is a DMBC. Then,  we have the upper bound
\begin{align}
\rvE_{\scC}&[I(\KA; Z^n,\Phi| \scC)]  \nnum\\
& \le  c(\alpha) \, |\calK|^{\alpha} |\alppmsg|^{\alpha}  |\calM|^{-\alpha}   \prod_{i=1}^n \sum_{s_i,x_i,z_i} \Upsilon(s_i,x_i,z_i,\alpha) \nnum\\
&    =  c(\alpha) \,|\calK|^{\alpha} |\alppmsg|^{\alpha}  |\calM|^{-\alpha}    \left[\sum_{s,x,z} \Upsilon( s,x,z,\alpha) \right]^n \label{eqn:multi-let5}   \ .
\end{align}
Note that the bound~\eqref{eqn:multi-let5}  holds for all $0<\alpha \le 1$. Recall also that $\calK =[1:2^{n\Rsk}]$, $\alppmsg = [1:2^{nR_{\Phi}}]$ and $\calM =  [1:2^{nR_M}]$ so $|\calK|^{\alpha} |\alppmsg|^{\alpha}  |\calM|^{-\alpha}  \doteq 2^{n \alpha(\Rsk+ R_{\Phi} - R_M)}$.  Now take the normalized logarithm and limit inferior of~\eqref{eqn:multi-let5} to get
\begin{align}
\liminf_{n\to\infty} & -\frac{1}{n}  \log \rvE_{\scC}[I(\KA; Z^n,\Phi| \scC)] \ge        \nnum\\
 & -\alpha(\Rsk+ R_{\Phi} - R_M)  -\log    \sum_{s,x,z}\Upsilon( s,x,z,\alpha)    \ . \nnum
\end{align}
The   joint distribution of $(X,Z,S)$, namely  $p(x,z,s) =   p(x,z|s)p(s)$, is induced by a particular input distribution $p(s)$. Essentially what we have done in this part of the proof is to develop a ``hybrid'' of the information leakage exponent for the wiretap channel model~\cite[Eq.\ (14)]{Hayashi} and the excited source model~\cite[Theorem~3]{chou_it10}.  Hence, an achievable exponent for the key leakage  given input distribution $p(s)$  is
$
F_{\mathrm{o}}(p(s),\Rsk, R_{\Phi},R_M  )$  defined in \eqref{eqn:secrecy_expo}.
The secrecy exponent part of the theorem  is proved for the random binning secret key code.   \\

\noindent {\em From Random Codes to a Deterministic Code}: Combining the proof in Section~\ref{sec:reliable_prf} and proof in this section, we have shown that for the $(2^{n\Rsk}, 2^{nR_M}, 2^{nR_{\Phi}},n)$   random binning secret key code, the expected   probability of error decays  with exponent (at least) $\Eo$ (expectation over codebooks and random binning functions) and the expected   key leakage decays exponentially with exponent (at least) $\Fo$.   Since both are measured with respect  the same (known) channel, there exists a binning secret key code  that meets the ensemble behavior.  More precisely, observe that $\rvP(\calE  ) = \rvE_{\scC}[\rvP(\calE  |\scC) ] =  \sum_{\calC}p(\calC) \rvP(\calE  |\scC=\calC)$, where $\calC$ runs through all binning secret key codes (a random code and two random binning functions) and the event $\calE$ is defined in \eqref{eqn:eventE}. By Markov's inequality,
\begin{equation}
\rvP_{\scC} \left[ \rvP(\calE  |\scC)\ge 3 \rvP(\calE   ) \right]\le\frac{1}{3} \  . \label{eqn:Pcode1}
\end{equation}
Similarly, when averaged over all codes, the average key leakage is $\rvE_{\scC}[I(\KA; Z^n,\Phi   |\scC) ]  =\sum_{\calC}  p(\calC)I(\KA; Z^n,\Phi |\scC=\calC)$, so by Markov's inequality,
\begin{equation}
\rvP_{\scC} \left[ I(\KA; Z^n,\Phi   |\scC)\ge 3 \rvE_{\scC}[I(\KA; Z^n,\Phi   |\scC) ] \right]\le\frac{1}{3} \  .  \label{eqn:Pcode2}
\end{equation}
From \eqref{eqn:Pcode1}, by considering the complement of the event of interest, we can conclude that there exists a subset of  binning secret key codes $\calD_1$ with total probability mass that exceeds $2/3$   (i.e., $\sum_{\calC\in\calD_1}p(\calC) \ge 2/3$) such that $\rvP(\calE  |\scC=\calC) <  3\rvP(\calE)$ for every $\calC\in \calD_1$. Similarly, from~\eqref{eqn:Pcode2} there exists a subset of  binning secret key codes $\calD_2$ with total probability mass that exceeds  $2/3$ (i.e., $\sum_{\calC\in\calD_2}p(\calC)\ge 2/3$) such that $ I(\KA; Z^n,\Phi |\scC\!=\!\calC) < 3\rvE_{\scC}[I(\KA; Z^n,\Phi|\scC)] $ for every $\calC\in\calD_2$.  Note that $\rvP(\calD_1\cap\calD_2)\ge 1/3$ so $\calD_1\cap\calD_2\ne \emptyset$. Thus, there exists at least one   binning secret key code  $\calC^*$  in the ensemble of (good) codes  $\calD_1\cap\calD_2$ such that $\rvP(\calE_{\mathrm{key}} |\scC=\calC^*)\le\rvP(\calE|\scC=\calC^*)\dotleq 2^{-n\Eo}$ and $I(\KA;Z^n,\Phi|\scC=\calC^*)\dotleq 2^{-n\Fo}$, where  the event  $\calE_{\mathrm{key}}$ is defined in~\eqref{eqn:eventK}. \qed\\

\noindent {\em Proof of Lemma~\ref{lem:info_leak}}:  Recall the assumption that  the key and public message binning processes are random, uniform and independent of the random codewords (See Section~\ref{sec:defs} for definitions and the code construction).    The key leakage can be expressed as follows:
\begin{align}
\rvE_{\scC} [ I&( \KA;  Z^n,\Phi | \scC) ] = \rvE_{\scC} [ H(\KA| \scC)-H(\KA|Z^n,\Phi| \scC)] \nnum \\
&= \rvE_{\scC} [H(\KA| \scC)+ H(\Phi|Z^n, \scC) -H(\KA,\Phi|Z^n, \scC)] \nnum \\
&\le \log |\calK|+\log |\alppmsg|  - \rvE_{\scC}[H(\KA,\Phi|Z^n, \scC)]  \ . \label{eqn:upper_bd_mi}
\end{align}
The conditioning is on the specific codebook used, i.e., $\scC=\calC$. It remains to lower bound the conditional entropy in~\eqref{eqn:upper_bd_mi}.   For this purpose, let
\begin{equation}
H_{1+\alpha}(X)\triangleq -\frac{1}{\alpha} \log \sum_{x\in\calX} p(x)^{1+\alpha} \label{eqn:renyi}
\end{equation}
be  the {\em R\'{e}nyi entropy} of order $1+\alpha$ for $0<\alpha\le 1$. Note that $\lim_{\alpha\searrow 0} H_{1+\alpha}(X) = H(X)$. Also, by the concavity of $t\mapsto\log t$, it can be verified that $H(X)\ge H_{1+\alpha}(X)$ for all $0<\alpha\le 1$.   Consider the conditional entropy  in~\eqref{eqn:upper_bd_mi},
\begin{align}
& \rvE_{\scC}[H(\KA,\Phi|Z^n, \scC)]  \nonumber \\
&=  \rvE_{\scC}\left[ \sum_{z^n} p(z^n ) H (\KA,\Phi|Z^n = z^n, \scC ) \right]  \nnum \\
&\ge  \sum_{z^n} p(z^n ) \rvE_{\scC}[ H_{1+\alpha}(\KA,\Phi|Z^n  = z^n, \scC ) ]  \label{eqn:renyi1} \\
&\ge \! \sum_{z^n}  p(z^n )\! \left( \! -\frac{1}{\alpha} \log \rvE_{\scC} \left[ \sum_{(\kA, \phi)\in\calK\times\alppmsg}\!\!\! p(\kA,\phi|z^n,\scC)^{1+\alpha}   \!\right] \right) \! .  \label{eqn:renyi_lb}
\end{align}
The last inequality is due to the definition of R\'{e}nyi entropy  in~\eqref{eqn:renyi} and the application of  Jensen's inequality noting that  the function $x\mapsto -\log x$ is convex.

Now let  $(\tilM,\tilX^n)$ be  a pair of random variables identically distributed to, but conditionally independent of $(M,X^n)$ given the events $\{Z^n=z^n\}$ and $\{\scC=\calC\}$.  Recall that $k(\fndot,\fndot)$ and $\phi(\fndot,\fndot)$ are  the key and public message  random binning functions respectively. See~\eqref{eqn:phi_def} and~\eqref{eqn:ka_def} for   definitions.   Define $(\tilde{K}_{\mathrm{A}},\tilde{\Phi}) \triangleq (k (\tilM,\tilX^n), \phi(\tilM,\tilX^n))$.  Then,
\begin{align}
& p(\kA,\phi |z^n, \calC)^{1+\alpha}     \nnum\\
&= p(\kA,\phi|z^n, \calC) \rvP \left[ (\tilde{K}_{\mathrm{A}},\tilde{\Phi})  = (\kA,\phi) | Z^n=z^n, \scC=\calC  \right]^{\alpha}   ,   \label{eqn:prob_bin}
\end{align}
by interpreting the R\'{e}nyi entropy  in~\eqref{eqn:renyi} in terms of an independent [from $({K}_{\mathrm{A}},{\Phi})$] and  identically distributed random variable $(\tilde{K}_{\mathrm{A}},\tilde{\Phi})$.

Define a shorthand notation for the indicator function as
\begin{align}
\indDD \triangleq {\bf 1}[k_{\calC}(m, x^n)=\kA ,  \phi_{\calC}(m, x^n)=\phi]. \label{eq.shorthand}
\end{align}
where $k_{\calC}(\fndot)$ and $\phi_{\calC}(\fndot)$ are the binning functions associated to a specific codebook $\scC=\calC$.  We upper bound the expectation in the  logarithm in \eqref{eqn:renyi_lb} on the top of  the next page.

\begin{figure*}
\begin{align}
& \rvE_{\scC}
    \Bigg\{
        \sum_{\kA,\phi}  p(\kA,\phi|z^n, \scC) \rvP \left[ (\tilde{K}_{\mathrm{A}},\tilde{\Phi})  = (\kA,\phi) | Z^n=z^n, \scC  \right]^{\alpha}
    \Bigg\}  \label{eqn:subst} \\
& = \rvE_{\scC}
     \Bigg\{
    \sum_{\kA, \phi}
    \Bigg[
        p(\kA,\phi|z^n, \scC) 
        \Bigg(\sum_{\kA', \phi'} p(k'_{\mathrm{A}},\phi'|z^n, \scC) \indA \Bigg)^{\alpha}
    \Bigg]
    \Bigg\} \label{eqn:write_pr} \\
& = \rvE_{\scC}
    \Bigg\{
    \sum_{\kA, \phi}
    \Bigg[
        \Bigg( \sum_{m,x^n} p(m,x^n|z^n) \indD \Bigg)
        \Bigg( \sum_{\kA', \phi'} p(k'_{\mathrm{A}},\phi'|z^n, \scC) \indA \Bigg)^{\alpha}
    \Bigg]
    \Bigg\} \label{eq.applyBayesA}   \\ 
& = \rvE_{\scC}
    \Bigg\{
        \sum_{m,x^n}  p(m, x^n|z^n) 
        \Bigg[
            \sum_{\kA, \phi} \indD
            \Bigg(\sum_{k'_{\mathrm{A}}, \phi'} p(k'_{\mathrm{A}}, \phi'|z^n,\scC) \indA
            \Bigg)^{\alpha}
       \Bigg]
    \Bigg\} \label{eq.reorder} \\ 
& \le \rvE_{\scC}
    \Bigg\{
        \sum_{m,x^n} p(m, x^n|z^n) 
        \Bigg[ \sum_{\kA, \phi} \indD
            \Bigg(\sum_{k'_{\mathrm{A}}, \phi'} p(k'_{\mathrm{A}}, \phi'|z^n,\scC) \indA
            \Bigg)
        \Bigg]^{\alpha}
    \Bigg\} \label{eq.jensensA} \\
& = \rvE_{\scC}
    \Bigg
    \{\sum_{m,x^n}  p(m,x^n|z^n) 
        \Bigg[\sum_{\kA, \phi} \indD \nnum\\
&\qquad\qquad \times        \Bigg(\sum_{k'_{\mathrm{A}}, \phi'}
            \Bigg(\sum_{m',x'^n} p(m',x'^n|z^n) \indB \Bigg)\indA
        \Bigg)
        \Bigg]^{\alpha}
      \Bigg\}  \label{eq.applyBayesB}  \\ 
& = \rvE_{\scC}
    \Bigg\{
        \sum_{m,x^n}  p(m,x^n|z^n) 
        \Bigg[ \sum_{m',x'^n} p(m',x'^n|z^n)\nnum\\
      &\qquad\qquad       \times \Bigg(\sum_{\kA, \phi} \sum_{k'_{\mathrm{A}}, \phi'} \indD \indB \indA \Bigg)
        \Bigg]^{\alpha}
    \Bigg\} \label{eqn:reoder}\\
    & = \rvE_{\scC}
    \Bigg\{
        \sum_{m,x^n}  p(m,x^n|z^n) 
        \Bigg[ \sum_{m',x'^n} p(m',x'^n|z^n)
            \Bigg(\sum_{\kA, \phi}  \indD  \mathbf{1} [ \kA,  \phi|m', x'^n, \scC]\Bigg)
        \Bigg]^{\alpha}
    \Bigg\} \label{eqn:sift}\\
& = \rvE_{\scC}
    \Bigg\{ \sum_{m,x^n}  p(m,x^n|z^n) 
        \Bigg[ p(m,x^n|z^n)  \nnum\\  
             &\qquad\qquad +    \sum_{(m',x'^n) \neq (m,x^n)} p(m',x'^n|z^n) 
            \Bigg(\sum_{\kA, \phi} \indD \indE \Bigg)
        \Bigg]^{\alpha}\Bigg\} \label{eq.sifting} \\
& \leq \rvE_{\scC}
    \Bigg\{
        \sum_{m,x^n} p(m,x^n|z^n) 
        \Bigg\{ p(m,x^n|z^n)^{\alpha}
             \nnum\\
           &\qquad\qquad  +  \Bigg[
            \sum_{(m',x'^n) \neq (m,x^n)} p(m',x'^n|z^n)  \Bigg(\sum_{\kA, \phi} \indD \indE \Bigg)
            \Bigg]^{\alpha}
        \Bigg\}
    \Bigg\} \label{eq.applyInequ} \\
& \leq \sum_{m,x^n} p(m,x^n|z^n)^{1 + \alpha}  \nnum\\
&\qquad\qquad  +
    \Bigg[
        \rvE_{\scC}
        \Bigg\{\sum_{m,x^n}  p(m,x^n|z^n) 
        \sum_{(m',x'^n)\neq (m,x^n)} p(m', x'^n|z^n)
            \Bigg(\sum_{\kA, \phi} \indD \indE \Bigg)
        \Bigg\}
    \Bigg]^{\alpha} \label{eq.jensensB} \\
& = \sum_{m,x^n} p(m,x^n|z^n)^{1 + \alpha}
    + \Bigg[ \sum_{m,x^n} p(m,x^n|z^n)
        \sum_{(m',x'^n)\neq (m,x^n)} p(m,x^n|z^n)
            \Bigg(\sum_{\kA, \phi} \frac{1}{(|{\calK}||{\alppmsg}|)^2}  \Bigg)
     \Bigg]^{\alpha} \label{eq.randomBinning} \\
& = \sum_{m,x^n} p(m,x^n|z^n)^{1 + \alpha} + \frac{1}{|\calK|^{\alpha}|\alppmsg|^{\alpha}}
       \Bigg[\sum_{m,x^n} \sum_{(m',x'^n)\neq (m,x^n)} p(m,x^n|z^n) p(m',x'^n|z^n) \Bigg]^{\alpha}
       \label{eqn:pull_out} \\
& \leq \sum_{m,x^n}  p(m,x^n|z^n)^{1 + \alpha} + \frac{1}{|\calK|^{\alpha}|\alppmsg|^{\alpha}}. \label{eq.unityUpperBnd}
\end{align}
\hrulefill
\vspace * {4pt}
\end{figure*}

The step \eqref{eqn:subst} is a result of plugging \eqref{eq.shorthand} into the argument of the logarithm in~\eqref{eqn:renyi_lb}. The step \eqref{eqn:write_pr} follows by writing out the probability of a collision event in~\eqref{eqn:prob_bin} explicitly as a sum. The step in \eqref{eq.applyBayesA} applies the law of total probability. We sum over all possible $(m, x^n)$ that are assigned bin indices $(\kA, \phi)$ for a given pair of binning function indexed by $\scC$. Equation~\eqref{eq.reorder}  follows by simple reordering of the sums.

The step \eqref{eq.jensensA} is
an application of Jensen's Inequality to the term in brackets
$[\fndot]^{\alpha}$ since the sum over $(\kA,\phi)$ is a sum over the
probability mass function ${\bf 1}[\kA, \phi|m, x^n, \calC]$
(cf.~(\ref{eq.shorthand}) for the definition of this indicator
function). Also, the function $x\mapsto x^{\alpha}$ is concave for $\alpha \in [0,1]$.   We recall that $m, x^n$, and $\calC$ are all
fixed for this inner sum, the last being fixed by the outer
expectation over $\scC$. Equation~\eqref{eq.applyBayesB} follows from the same reasoning as~\eqref{eq.applyBayesA}, i.e., the law of total probability. Equation~\eqref{eqn:reoder} follows by simple reordering of the sums.

In \eqref{eqn:sift}, we used the ``sifting'' property of the indicator function  $\indA$.
In \eqref{eq.sifting} we split the sum over $(m',x'^n)$ into two terms
and distributed the sums over $(k_{\mathrm{A}}' , \phi')$. Note that for the $(m', x'^n) = (m, x^n)$ term,  $\sum_{\kA, \phi} \mathbf{1}[\kA, \phi | m, x^n, \scC]=1$.  We next applied the inequality $(x+y)^{\alpha} \leq x^{\alpha} +
y^{\alpha}$, for $0 \leq \alpha \leq 1$ to get \eqref{eq.applyInequ}.

In \eqref{eq.jensensB} we note that the first term is not a function
of $\calC$. Using the concavity of $x\mapsto x^{\alpha}$ (for $\alpha \in [0,1]$), we move both the sum
over $(m,x^n)$ and the expectation over codebooks inside the function, a
step justified by Jensen's Inequality.

In~(\ref{eq.randomBinning}) we apply the uniformly random design of
the binning functions.  Since $(m,x^n) \neq (m',x'^n)$ {\em for every} term in
the sum, each of the indicator functions equals the (fixed) pair $(\kA,
\phi)$ with equal probability and independently.  Thus, the
probability that {\em both} equal $(\kA, \phi)$ is the square (by the
independence) of the reciprocal of the number of possibilities (by the
uniformity), i.e., $\rvE_{\scC} [ \indD \indE ] = (|\calK||\alppmsg|)^{-2}$. In \eqref{eqn:pull_out}, we pulled out $(|\calK||\alppmsg|)^{-\alpha}$. Finally, we note  that $p(m,x^n|z^n) p(m', x'^n|z^n)$
is a well defined (conditional) pmf and that we
are missing one term in the double sum. Hence,  we get  \eqref{eq.unityUpperBnd}  by upper bounding the double sum by one.

Substituting~\eqref{eq.unityUpperBnd} back into~\eqref{eqn:renyi_lb}    gives
\begin{align}
& \rvE_{\scC}[ H(\KA,\Phi|Z^n, \scC)]  \nnum\\
&\ge  \sum_{z^n}\!  p(z^n ) \! \left[  -\frac{1}{\alpha} \log \left( \frac{1}{|\calK|^{\alpha}|\alppmsg|^{\alpha}}  \! +\! \sum_{m,x^n } p(m,x^n|z^n)^{1+\alpha} \right)\right]  \nnum \\
&=  \log( |\calK||\alppmsg|) -\frac{1}{\alpha} \sum_{z^n} p(z^n)  \times\ldots\nnum\\
&\quad\qquad\times  \log \left( 1+  |\calK|^{\alpha}|\alppmsg|^{\alpha}  \sum_{m,x^n}  p(m,x^n|z^n)^{1+\alpha}  \right )  \label{eqn:lemma_end1} \\
&\ge  \log( |\calK||\alppmsg|)   - \left(\frac{\log e}{\alpha} \right)|\calK|^{\alpha}|\alppmsg|^{\alpha}    \times\ldots\nnum\\
&\quad\qquad\times  \sum_{z^n} p(z^n ) \sum_{m,x^n } p(m,x^n|z^n)^{1+\alpha}  \ , \label{eqn:lemma_end}
\end{align}
where in \eqref{eqn:lemma_end1} we pulled out the $|\calK|^{-\alpha}|\alppmsg|^{-\alpha}$  term from the logarithm above and  in~\eqref{eqn:lemma_end} we applied the relation $\log(1+t)\le t \log e$ (recall that $\log=\log_2$). The proof of the lemma is completed by uniting~\eqref{eqn:upper_bd_mi} and~\eqref{eqn:lemma_end}.

\subsection*{Acknowledgments}
The authors would like to acknowledge one of the reviewers whose
insights led to the discussion on the connection of our work to that
in Csisz\'{a}r and Narayan \cite{CsiszarN04} and Gohari and
Anantharam~\cite{gohari_I} in Section~\ref{sec:inner}.

{\footnotesize
\bibliographystyle{ieeetr}
\bibliography{references} }

\begin{thebibliography}{10}

\bibitem{chou_11}
T.-H. Chou, V.~Y.~F. Tan, and S.~C. Draper, ``On the capacity of the
  sender-excited secret key agreement model,'' in {\em Proc. Allerton
  Conference on Communication, Control, and Computing}, 2011.

\bibitem{Liang}
Y.~Liang, H.~V. Poor, and S.~Shamai, {\em Information Theoretic Security}.
\newblock Now Publishers Inc, 2009.

\bibitem{Ahlswede_Csiszar93}
R.~Ahlswede and I.~Csisz\'{a}r, ``Common randomness in information theory and
  cryptography part {I}: Secret sharing,'' {\em IEEE Trans. Inform. Theory},
  vol.~39, no.~4, pp.~1121--1132, 1993.

\bibitem{maurer93}
U.~M. Maurer, ``Secret key agreement by public discussion from common
  information,'' {\em IEEE Trans. Inform. Theory}, vol.~39, no.~3,
  pp.~733--742, 1993.

\bibitem{Weissman10}
T.~Weissman, ``Capacity of channels with action-dependent states,'' {\em IEEE
  Trans. Inform. Theory}, vol.~56, pp.~5396--5411, Nov 2010.

\bibitem{Asnani_ProbingCapacity}
H.~{Asnani}, H.~{Permuter}, and T.~{Weissman}, ``{Probing Capacity},'' {\em
  IEEE Trans. Inform. Theory}, vol.~57, pp.~7317--7332, Nov 2011.

\bibitem{Kit10}
K.~Kittichokechai, T.~J. Oechtering, M.~Skoglund, and R.~Thobaben, ``Source and
  channel coding with action-dependent partially known two-sided state
  information,'' in {\em Proc. Int. Symp. Inform. Theory}, pp.~629--633, June
  2010.

\bibitem{Per11}
H.~Permuter and T.~Weissman, ``Source coding with a side information ``vending
  machine'','' {\em IEEE Trans. Inform. Theory}, vol.~57, pp.~4530--4544, Jul
  2011.

\bibitem{csiszar2000}
I.~Csisz\'{a}r and P.~Narayan, ``Common randomness and secret key generation
  with a helper,'' {\em IEEE Trans. Inform. Theory}, vol.~46, no.~2,
  pp.~344--366, 2000.

\bibitem{gallagerIT}
R.~G. Gallager, {\em Information theory and reliable communication}.
\newblock New York: Wiley, 1968.

\bibitem{gallager76}
R.~G. Gallager, ``Source coding with side information and universal coding,''
  {\em M.I.T. LIDS-P-937}, 1976.

\bibitem{Hayashi}
M.~Hayashi, ``Exponential decreasing rate of leaked information in universal
  random privacy amplification,'' {\em IEEE Trans. Inform. Theory}, vol.~57,
  pp.~3989--4001, June 2011.

\bibitem{Khisti_isit08}
A.~Khisti, S.~Diggavi, and G.~Wornell, ``Secret-key generation with correlated
  sources and noisy channels,'' in {\em Proc. Int. Symp. Inform. Theory},
  pp.~1005--1009, July 2008.

\bibitem{Prabhakaran08}
V.~Prabhakaran, K.~Eswaran, and K.~Ramchandran, ``Secrecy via sources and
  channels -- a secret key-secret message rate tradeoff region,'' in {\em Proc.
  Int. Symp. Inform. Theory}, pp.~1010--1014, July 2008.

\bibitem{CsiszarN04}
I.~Csisz\'{a}r and P.~Narayan, ``The secret key capacity of multiple
  terminals,'' {\em IEEE Trans. Inform. Theory}, vol.~50, pp.~3047--3061, Dec
  2004.

\bibitem{CsiszarN08}
I.~Csisz\'{a}r and P.~Narayan, ``Secrecy capacities for multiterminal channel
  models,'' {\em IEEE Trans. Inform. Theory}, vol.~54, pp.~2437--2452, Jun
  2008.

\bibitem{ChenVinck06}
Y.~Chen and A.~J. Han~Vinck, ``Wiretap channel with side information,'' {\em
  IEEE Trans. Inform. Theory}, vol.~54, pp.~395--402, Jan. 2008.

\bibitem{Liu07}
W.~Liu and B.~Chen, ``Wiretap channel with two-sided channel state
  information,'' in {\em Proc. Asilomar Conf. Signals, Systems and Computers,
  2007}, pp.~893 --897, Nov. 2007.

\bibitem{Chia10}
Y.~K. {Chia} and A.~{El Gamal}, ``{Wiretap channel with causal state
  information},'' {\em IEEE Trans. Inform. Theory}, vol.~58, pp.~2838--2849,
  May 2012.

\bibitem{Khisti_isit09}
A.~Khisti, S.~Diggavi, and G.~Wornell, ``Secret key agreement using asymmetry
  in channel state knowledge,'' in {\em Proc. Int. Symp. Inform. Theory},
  pp.~2286--2290, 2009.

\bibitem{Khisti11}
A.~Khisti, S.~Diggavi, and G.~Wornell, ``Secret-key agreement with channel
  state information at the transmitter,'' {\em IEEE Trans.\ on Foren.\ and
  Sec.}, vol.~6, pp.~672--681, Sep 2011.

\bibitem{chou_it10}
T.~Chou, S.~C. Draper, and A.~Sayeed, ``Key generation using external source
  excitation: Capacity, reliability, and secrecy exponent,'' {\em IEEE Trans.
  Inform. Theory}, vol.~58, pp.~2455--2474, Apr. 2012.

\bibitem{WTS07}
R.~Wilson, D.~Tse, and R.~A. Scholtz, ``Channel identification: Secret sharing
  using reciprocity in ultrawideband channels,'' {\em IEEE Trans. Inform.
  Foren. and Sec.}, vol.~2, pp.~364--375, Sep. 2007.

\bibitem{ARKA11}
A.~Agrawal, Z.~Rezki, A.~Khisti, and M.~Alouini, ``Noncoherent capacity of
  secret-key agreement with public discussion,'' {\em IEEE Trans. Inform.
  Foren. and Sec.}, vol.~6, pp.~565--574, Sept. 2011.

\bibitem{Hayashi06}
M.~Hayashi, ``General nonasymptotic and asymptotic formulas in channel
  resolvability and identification capacity and their application to the
  wiretap channel,'' {\em IEEE Trans. Inform. Theory}, vol.~52, pp.~1562--1575,
  April 2006.

\bibitem{Han}
T.~S. Han, {\em Information-Spectrum Methods in Information Theory}.
\newblock Springer, 2002.

\bibitem{Bloch}
M.~{Bloch} and J.~N. {Laneman}, ``{Secrecy from Resolvability},'' {\em
  arXiv:1105.5419}, May 2011.

\bibitem{BBCM95}
C.~Bennett, G.~Brassard, C.~Crepeau, and U.~Maurer, ``Generalized privacy
  amplification,'' {\em IEEE Trans. Inform. Theory}, vol.~41, pp.~1915--1923,
  Nov 1995.

\bibitem{Wat10}
S.~Watanabe, R.~Matsumoto, and T.~Uyematsu, ``{Strongly Secure Privacy
  Amplification Cannot Be Obtained by Encoder of {Slepian-Wolf} Code},'' {\em
  IEICE Transactions on Fundamentals of Electronics, Communications and
  Computer Sciences}, vol.~E93.A, no.~9, pp.~1650--1659, 2010.

\bibitem{ElGamal_Kim_LNIT}
A.~{El Gamal} and Y.-H. {Kim}, {\em Network Information Theory}.
\newblock Cambridge University Press, 2012.

\bibitem{Mau00}
U.~Maurer and S.~Wolf, ``Information-theoretic key agreement: From weak to
  strong secrecy for free,'' in {\em Lecture Notes in Computer Science},
  pp.~351--368, Springer-Verlag, 2000.

\bibitem{Maurer94}
U.~M. Maurer, ``The strong secret key rate of discrete random triples,'' {\em
  Communications and Cryptography: Two Sides of One Tapestry}, pp.~271--285,
  Nov 1994.

\bibitem{Wyner75}
A.~D. Wyner, ``The wire-tap channel,'' {\em The Bell Systems Technical
  Journal}, vol.~54, pp.~1355--1387, 1975.

\bibitem{gohari_I}
A.~A. Gohari and V.~Anantharam, ``Information-theoretic key agreement of
  multiple terminals -- {I}: Source model,'' {\em IEEE Trans. Inform. Theory},
  vol.~56, pp.~3973--3996, Aug 2008.

\bibitem{gohari_II}
A.~A. Gohari and V.~Anantharam, ``{Information-Theoretic Key Agreement of
  Multiple Terminals--Part II: Channel Model },'' {\em IEEE Trans. Inform.
  Theory}, vol.~56, pp.~3997--4010, Aug. 2010.

\bibitem{SlepianWolf}
D.~Slepian and J.~Wolf, ``Noiseless coding of correlated sources,'' {\em IEEE
  Trans. Inform. Theory}, vol.~19, pp.~471--480, Jul 1973.

\bibitem{TK12}
V.~Y.~F. Tan and O.~Kosut, ``On the dispersions of three network information
  theory problems,'' {\em arXiv:1201.3901}, Feb 2012.
\newblock [Online].

\end{thebibliography}

\end{document}